\documentclass[onecolumn,showkeys]{openjournal}
\usepackage{amsmath}
\def \be{\begin{equation}}
\def \ee{\end{equation}}
\def \bdm{\begin{eqnarray}}
\def \edm{\end{eqnarray}}
\begin{document}
\title{Boris and Exponential Integrators in the Theory of Particles Interacting with Magnetic Turbulence}
\author{A. Shalchi}
\affiliation{Department of Physics and Astronomy, University of Manitoba, Winnipeg, Manitoba R3T 2N2, Canada}
\email{andreasm4@yahoo.com}
\begin{abstract}
The interaction of electrically charged particles with magnetic fields is a fundamental problem in several areas of physics.
An example is the motion of energetic particles through a magnetized plasma. The most accurate and reliable way to explore
theoretically the interactions between particles and fields is via test-particle simulations. In such simulations one
creates the turbulent magnetic field and solves the Newton-Lorentz equation numerically by employing an integration scheme.
In the current article we discuss exponential integrators and derive systematically from this the Rodrigues scheme
as well as the famous Boris integrator. For an approach where one creates the magnetic field anew at each time step,
both integrators are overall comparable. In theory the Rodrigues approach should be more accurate due to the fact that
the occurring matrix exponential is evaluated without further approximations. Practically, both methods provide very
similar results. It is argued in the current article that a Rodrigues based integrator is a very strong alternative
because for the specific problem discussed here, it does not require longer computing times.
\end{abstract}
\section{Introduction}
The motion of electrically charged energetic particles in a plasma is difficult to describe analytically due to complicated
interactions between the particles and turbulent magnetic fields. However, this problem is fundamental and the 
knowledge of particle diffusion coefficients is important in several areas of research. The most prominent examples
are studies of solar modulation and diffusive shock acceleration (see, e.g., Zank et al. (2000), Li et al. (2003),
Li et al. (2005), Zank et al. (2006), Dosch \& Shalchi (2010), Li et al. (2012), Ferrand et al. (2014), Hu et al. (2017),
Qin et al. (2018), Shen \& Qin (2018), Moloto \& Engelbrecht (2020), Engelbrecht \& Wolmarans (2020), Engelbrecht \& Moloto (2021),
Shen et al. (2021), and Ngobeni et al. (2022)).

Ideally, one describes the motion of energetic particles through a plasma by using analytical theory. However, this task
is difficult due to the nonlinearity of the problem (see Shalchi (2009), Shalchi (2020), and Effenberger et al. (2025)
for reviews). Especially in analytical treatments of the transport, one has to distinguish between transport along and
across the mean magnetic field. Parallel diffusion is controlled by pitch-angle scattering and, associated with this,
a pitch-angle isotropization process. Perpendicular transport, on the other hand, depends on the structure of magnetic
field lines because initially particles follow individual lines. As soon as the transverse complexity of the turbulence
becomes manifested, particles leave their original field lines and the transport become diffusive. In the past it
was often assumed that perpendicular transport is more difficult to describe analytically, but significant progress
has been achieved during the past two decades (see, e.g., Matthaeus et al. (2003), Shalchi (2010), and Shalchi (2021)).

An alternative to analytical theories is to perform test-particle simulations. In such simulations the turbulence
is created artificially by using a certain spectrum in combination with random numbers. For a given turbulence model,
particle orbits can then be obtained by solving the Newton-Lorentz equation numerically. Considering and averaging
thousands of particle trajectories allows one to determine the diffusion coefficients in the different directions
of space. The advantage of simulations compared to analytical theories is obvious: while theories require to
make several assumptions which are often difficult to justify, the simulations are based on an exact (besides
numerical inaccuracies) particle orbits. A disadvantage of the simulations is, of course, that they cannot be
implemented in applications such as diffusive shock acceleration.

In the past, test-particle simulations have been performed by numerous authors in the context of space physics and
astrophysics (see, e.g., Qin et al. (2002a), Qin et al. (2002b), Qin et al. (2006), Zimbardo et al. (2006),
Perri \& Zimbardo (2007), Pommois et al. (2007), Perri \& Zimbardo (2009a), Perri \& Zimbardo (2009b), Tautz (2010),
Zimbardo et al. (2012), Hussein et al. (2015), Hussein \& Shalchi (2016), Heusen \& Shalchi (2016), Ivascenko et al. (2016),
Reichherzer et al. (2022), Snodin et al. (2022), and Els \& Engelbrecht (2024)). In the aforementioned simulations
different methods have been applied. For instance, there are grid-based simulations in which the turbulent magnetic
field is computed over a whole grid and thereafter the particle orbits are determined. In other simulation codes
the magnetic field is only computed at the current particle position. Both methods have advantages and disadvantages
is terms of memory needs and run times. In either case, one has to solve the Newton-Lorentz equation numerically.

It became clear over years that one of the best possible integration schemes to solve the equations of motion is the
so-called \textit{Boris integrator} (see Boris (1970)). In the current article we review this type of integration scheme
and derive it systematically from exponential integrators. As alternative we explore the combination of such exponential
integrators with the so-called \textit{Rodrigues formula} (see Rodrigues (1840)). We compare both integration schemes
with the aim to see which one is better suited for test-particle simulations. By doing this we focus on the very
specific case of a purely magnetic scenario where particles interact with a constant mean magnetic field and a turbulent
component. Furthermore, we consider simulations in which the magnetic field is computed anew at each particle position.
We shall demonstrate that, for this special case, both integrator schemes, Boris as well as Rodrigues, are strong tools
with a slight advantage of the latter approach due to more accuracy in the phase. A comprehensive comparison of relativistic
particle integrators can be found in Ripperda et al. (2018). However, in the current article we focus on the very specific
problem of particles interacting with a purely magnetic field which consists of a constant mean field and a turbulence
component. Furthermore, the turbulent component is created in specific way as described later in this article.

\section{Equations of Motion}

The fundamental equation describing the motion of an electrically charged particle in a purely magnetic system is the
\textit{Newton-Lorentz equation}
\be
\frac{d \vec{p}}{d t} = \frac{q}{c} \vec{v} \times \vec{B} (\vec{x})
\label{NewtonLEq}
\ee
where we have used the relativistic momentum $\vec{p}=m \gamma \vec{v}$. In the latter relation, as well as in Eq. (\ref{NewtonLEq}),
we have used the particle charge $q$, the vacuum speed of light $c$, the particle velocity $\vec{v}$, the particle's rest mass $m$,
and the Lorentz factor $\gamma$. Furthermore, we assume that the total magnetic field is given by $\vec{B} (\vec{x}) = \delta \vec{B} (\vec{x}) + B_0 \hat{z}$
where $\delta \vec{B} (\vec{x})$ describes the magnetic fluctuations (the turbulence) and $B_0$ is the constant mean field pointing in the
$z$-direction. Note, here we have used \textit{cgs units} rather than \textit{SI units}. The three components of Eq. (\ref{NewtonLEq})
can easily be determined and one gets
\bdm
\dot{v}_x & = & \Omega v_y + \Omega \left( v_y \frac{\delta B_z}{B_0} - v_z \frac{\delta B_y}{B_0} \right), \nonumber\\
\dot{v}_y & = & - \Omega v_x + \Omega \left( v_z \frac{\delta B_x}{B_0} - v_x \frac{\delta B_z}{B_0} \right), \nonumber\\
\dot{v}_z & = & \Omega \left( v_x \frac{\delta B_y}{B_0} - v_y \frac{\delta B_x}{B_0} \right).
\label{threecomp}
\edm
In the latter equations of motion we have used the gyro-frequency
\be
\Omega = \frac{q B_0}{m c \gamma}
\label{defomega}
\ee
which is the angular velocity of the rotation if there was only a mean magnetic field $B_0$ and no turbulence. Therefore, we also call $\Omega$
the \textit{unperturbed gyro-frequency}. Note, for negative $q$ the gyro-frequency defined here becomes negative. This simply means that the particle
reverses its direction of rotation. The three components listed via Eq. (\ref{threecomp}) can be written in the form
\be
\frac{d}{d t} \vec{v} = \boldsymbol{M} \vec{v}
\label{diffeq1}
\ee
where we have used the $3 \times 3$ matrix
\bdm
\boldsymbol{M} = \frac{\Omega}{B_0} \left(
\begin{array}{ccc}
0      		\quad\quad & B_z		\quad\quad		& - B_y   	\\[0.2cm]
- B_z	    \quad\quad & 0 			\quad\quad		& B_x    	\\[0.2cm]
B_y			\quad\quad & - B_x		\quad\quad 		& 0
\end{array}
\right)
\label{thematrixMwithdBz}
\edm
where, in our case, $B_x = \delta B_x$, $B_y = \delta B_y$, and $B_z = B_0 + \delta B_z$. Eq. (\ref{diffeq1}) can formally be solved via
\be
\vec{v} \left( t \right) = e^{\int_{t_0}^t d \tau \; \boldsymbol{M} (\tau)} \vec{v} \left( t_0 \right).
\label{exactvelo}
\ee
Problematic here is that the matrix $\boldsymbol{M}$ does depend on time and, therefore, the matrix exponential is not trivial.
According to Eq. (\ref{thematrixMwithdBz}), the entries of the matrix $\boldsymbol{M}$ are the components of the magnetic field.
In general we have $\vec{B} = \vec{B} \left[ \vec{x} \left( t \right), t \right]$ because we trace energetic particles.
Therefore, the magnetic field has to be evaluated along the particle trajectory $\vec{x} \left( t \right)$. Furthermore,
the magnetic field can depend on time. In the current article, however, we only consider a static scenario meaning that
$\vec{B} = \vec{B} \left[ \vec{x} \left( t \right) \right]$. Test-particle simulations have been performed for
dynamical turbulence in Hussein \& Shalchi (2016) as well as in Heusen \& Shalchi (2016).

In the following we discuss two different numerical/computational schemes based on Eq. (\ref{exactvelo}).

\section{Matrix Exponentials and the Rodrigues Formula}\label{RodriguesSect}

Eq. (\ref{exactvelo}) would allow us to determine the particle velocity $\vec{v}$ at time $t$. However, this equation contains
an exponential where the argument is a time-integral over the matrix $\boldsymbol{M}$. Alternatively, we can write Eq. (\ref{exactvelo}) as
\be
\vec{v} \left( t + \Delta t \right) = e^{\int_{t}^{t + \Delta t} d \tau \; \boldsymbol{M} (\tau)} \vec{v} \left( t \right)
\label{exactvelo2}
\ee
where $\Delta t$ corresponds to the performed time step. As an approximation we assume that $\boldsymbol{M} (\tau)$ remains constant during
the time-interval $t \leq \tau \leq t + \Delta t$ so that
\be
\vec{v} \left( t + \Delta t \right) \approx e^{\boldsymbol{M} \Delta t} \vec{v} \left( t \right).
\label{vattplusdeltat}
\ee
In numerical schemes this becomes
\be
\vec{v}_{n+1} \approx e^{\boldsymbol{M} \Delta t} \vec{v}_n
\label{NumMatrixEq}
\ee
where $\Delta t$ is the (constant) time step size. Approximating the matrix in the exponential by a constant is called the
\textit{frozen field approximation}. Within this approach, the matrix $\boldsymbol{M}$ has to be evaluated at some time between
$t$ and $t + \Delta t$. One option is to use
\be
\boldsymbol{M} = \boldsymbol{M} \left( \vec{x}_n + \frac{1}{2} \vec{v}_n \Delta t \right)
\label{MinRodrigues}
\ee
meaning that we evaluate the magnetic field at the \textit{predicted midpoint}
\be
\vec{x} = \vec{x}_n + \frac{1}{2} \vec{v}_n \Delta t.
\label{predictedmidpoint}
\ee
The approach described here is based on \textit{midpoint rule}. It is well-known that employing this rule leads to a result which is correct
up to second order in the step size $\Delta t$ (see Burden \& Faires (2011)).

However, Eq. (\ref{NumMatrixEq}) still contains a matrix exponential but the matrix $\boldsymbol{M}$ is now a constant. A powerful tool to evaluate
this is provided by \textit{Rodrigues' rotation formula} which is derived in the following. To use this formula, it is convenient to define
\be
\omega := \Omega \frac{|\vec{B}|}{B_0} = \frac{q}{m c \gamma} |\vec{B}|
\label{definelittleomega}
\ee
which can be understood as instantaneous gyro-frequency in the perturbed field. Furthermore, we need to discuss the properties of the matrix $\boldsymbol{M}$
defined via Eq. (\ref{thematrixMwithdBz}). For the square of this matrix, for instance, one can easily derive
\bdm
\boldsymbol{M}^2 = \frac{\Omega^2}{B_0^2} \left(
\begin{array}{ccc}
-B_y^2 -B_z^2  	\quad\quad & B_x B_y		\quad\quad		& B_x B_z   	\\[0.2cm]
B_x B_y	    	\quad\quad & -B_x^2 -B_z^2 	\quad\quad		& B_y B_z   	\\[0.2cm]
B_x B_z			\quad\quad & B_y B_z		\quad\quad 		& -B_x^2 -B_y^2
\end{array}
\right).
\label{thematrixMwithdBz2}
\edm
This can be written in the more compact form
\be
\boldsymbol{M}^2 = \frac{\Omega^2}{B_0^2} \left[ \vec{B} \otimes \vec{B} - \vec{B}^2 \boldsymbol{I} \right]
\equiv \omega^2 \left[ \hat{B} \otimes \hat{B} - \boldsymbol{I} \right]
\ee
where we have used the unit vector
\bdm
\hat{B} = \frac{1}{\sqrt{B_x^2 + B_y^2 + B_z^2}}  \left(
\begin{array}{c}
B_x		   	\\[0.2cm]
B_y	       	\\[0.2cm]
B_z			
\end{array}
\right)
\edm
and the \textit{tensor product} $\otimes$ which is also known as \textit{dyadic} or \textit{outer product}. Furthermore, we have used the unit matrix $\boldsymbol{I}$.
One can show via direct calculation that the matrix $\boldsymbol{M}$ satisfies the following relations
\bdm
\boldsymbol{M}^3 & = & - \omega^2 \boldsymbol{M}, \nonumber\\
\boldsymbol{M}^4 & = & - \omega^2 \boldsymbol{M}^2, \nonumber\\
\boldsymbol{M}^5 & = & \omega^4 \boldsymbol{M}, \nonumber\\
\boldsymbol{M}^6 & = & \omega^4 \boldsymbol{M}^2.
\label{Mrelations}
\edm
We can generalize this so that for integer $n$ we have
\be
\boldsymbol{M}^{2n} = - \left( -1 \right)^n \omega^{2n-2} \boldsymbol{M}^2 \quad\textnormal{for}\quad n=1,2,3,\dots
\label{M2n}
\ee
and
\be
\boldsymbol{M}^{2n+1} = \left( -1 \right)^n \omega^{2n} \boldsymbol{M} \quad\textnormal{for}\quad n=0,1,2,\dots.
\label{M2n+1}
\ee
We conclude that any power of $\boldsymbol{M}$ can be expressed by either $\boldsymbol{M}$ or $\boldsymbol{M}^2$. For the matrix exponential
needed in Eq. (\ref{NumMatrixEq}) we can employ the Talyor-expansion
\be
e^{\boldsymbol{M} \Delta t} = \sum_{n=0}^{\infty} \frac{1}{n!} \left( \boldsymbol{M} \Delta t \right)^n.
\ee
The sum can be split into odd and even terms so that one gets
\be
e^{\boldsymbol{M} \Delta t} = \boldsymbol{I} + \sum_{n=0}^{\infty} \frac{1}{(2n+1)!} \boldsymbol{M}^{2n+1} \left( \Delta t \right)^{2n+1}
+ \sum_{n=1}^{\infty} \frac{1}{(2n)!} \boldsymbol{M}^{2n} \left( \Delta t \right)^{2n}.
\ee
Therein we can now employ Eqs. (\ref{M2n}) and (\ref{M2n+1}) yielding
\be
e^{\boldsymbol{M} \Delta t} = \boldsymbol{I} + \sum_{n=0}^{\infty} \frac{1}{(2n+1)!} \left( -1 \right)^n \omega^{2n} \left( \Delta t \right)^{2n+1} \boldsymbol{M}
- \sum_{n=1}^{\infty} \frac{1}{(2n)!} \left( -1 \right)^n \omega^{2n-2} \left( \Delta t \right)^{2n} \boldsymbol{M}^2.
\ee
We can further rewrite this via
\be
e^{\boldsymbol{M} \Delta t} = \boldsymbol{I} + \frac{1}{\omega} \sum_{n=0}^{\infty} \frac{1}{(2n+1)!} \left( -1 \right)^n \left( \omega \Delta t \right)^{2n+1} \boldsymbol{M}
+ \frac{1}{\omega^2} \boldsymbol{M}^2 - \frac{1}{\omega^2} \sum_{n=0}^{\infty} \frac{1}{(2n)!} \left( -1 \right)^n \left( \omega \Delta t \right)^{2n} \boldsymbol{M}^2.
\label{ExpintermsofMandM2}
\ee
Therein we can now see the occurrence of Taylor expansions of trigonometric functions. Thus, we can write
\be
e^{\boldsymbol{M} \Delta t} = \boldsymbol{I} + \frac{\sin \left( \omega \Delta t \right)}{\omega} \boldsymbol{M}
+ \frac{1 - \cos \left( \omega \Delta t \right)}{\omega^2} \boldsymbol{M}^2
\label{RodriguesFormula}
\ee
which is known as the \textit{Rodrigues rotation formula} (see Rodrigues (1840)). The latter formula is well-known and often used in
physics (see, e.g., Marsden \& Ratiu (1999)) and numerical integration methods (see Hairer et al. (2006)).

As discussed above, the magnetic field components are $B_x = \delta B_x$, $B_y = \delta B_y$, and $B_z = B_0 + \delta B_z$. For this case the matrix given
by Eq. (\ref{thematrixMwithdBz}) becomes
\bdm
\boldsymbol{M} = \Omega \left(
\begin{array}{ccc}
0      						\quad\quad & 1 + \delta B_z / B_0	\quad\quad & - \delta B_y / B_0   \\[0.2cm]
- 1 - \delta B_z / B_0    	\quad\quad & 0 						\quad\quad & \delta B_x / B_0     \\[0.2cm]
\delta B_y / B_0			\quad\quad & - \delta B_x / B_0		\quad\quad & 0
\end{array}
\right).
\label{thematrixMwithturbandmean}
\edm
Magnetic fluctuations in interplanetary and interstellar spaces can be measured \textit{in situ} via magnetometers aboard space probes such as \textit{Voyager 1}.
Zank et al. (2019) have shown based on such measurements that the solar wind and the very local interstellar medium are often nearly incompressible meaning that
$\delta B_z \approx 0$. Therefore, we now consider the nearly incompressible case for which our matrix (\ref{thematrixMwithturbandmean}) simplifies to
\bdm
\boldsymbol{M} = \Omega \left(
\begin{array}{ccc}
0      				\quad\quad & 1 					\quad\quad & - \delta B_y / B_0   \\[0.2cm]
- 1 			    \quad\quad & 0 					\quad\quad & \delta B_x / B_0     \\[0.2cm]
\delta B_y / B_0	\quad\quad & - \delta B_x / B_0	\quad\quad & 0
\end{array}
\right).
\label{thematrixM}
\edm
Note, there a scenarios where one finds significant compressibility in the heliosheath and the very local interstellar medium (see Zhao et al. (2020),
Fraternale et al. (2022), and Zhao et al. (2024)). The methods discussed throughout the current paper can easily be generalized to cover the incompressible
case where $\delta B_z \neq 0$. In the remainder of this article, however, we focus on the incompressible case and work with $\delta B_z = 0$.

Using Eq. (\ref{RodriguesFormula}) in Eq. (\ref{NumMatrixEq}) gives us for the components of the updated velocity
\bdm
v_{x,n+1} & = & v_{x,n} + \frac{\sin \left( \omega \Delta t \right)}{\omega/\Omega} \left[ v_{y,n} - \frac{\delta B_y}{B_0} v_{z,n} \right] \nonumber\\
& + & \frac{1 - \cos \left( \omega \Delta t \right)}{\omega^2/\Omega^2} \left[ - \left( 1 + \frac{\delta B_y^2}{B_0^2} \right) v_{x,n}
+ \frac{\delta B_x}{B_0} \frac{\delta B_y}{B_0} v_{y,n} + \frac{\delta B_x}{B_0} v_{z,n} \right], \nonumber\\
v_{y,n+1} & = & v_{y,n} + \frac{\sin \left( \omega \Delta t \right)}{\omega/\Omega} \left[ - v_{x,n} + \frac{\delta B_x}{B_0} v_{z,n} \right] \nonumber\\
& + & \frac{1 - \cos \left( \omega \Delta t \right)}{\omega^2/\Omega^2} \left[ \frac{\delta B_x}{B_0} \frac{\delta B_y}{B_0} v_{x,n}
- \left( 1 + \frac{\delta B_x^2}{B_0^2} \right) v_{y,n} + \frac{\delta B_y}{B_0} v_{z,n} \right], \nonumber\\
v_{z,n+1} & = & v_{z,n} + \frac{\sin \left( \omega \Delta t \right)}{\omega/\Omega} \left[ \frac{\delta B_y}{B_0} v_{x,n} - \frac{\delta B_x}{B_0} v_{y,n} \right] \nonumber\\
& + & \frac{1 - \cos \left( \omega \Delta t \right)}{\omega^2/\Omega^2} \left[ \frac{\delta B_x}{B_0} v_{x,n} + \frac{\delta B_y}{B_0} v_{y,n}
- \left( \frac{\delta B_x^2}{B_0^2} + \frac{\delta B_y^2}{B_0^2} \right) v_{z,n} \right]
\label{explicitvelocityupdate}
\edm
where we have also used Eqs. (\ref{thematrixMwithdBz2}) and (\ref{thematrixM}) to replace the matrices $\boldsymbol{M}$ and $\boldsymbol{M}^2$,
respectively. It needs to be emphasized that the magnetic field components in Eq. (\ref{explicitvelocityupdate}) have to be evaluated
at the predicted midpoint given by Eq. (\ref{predictedmidpoint}). Note, sometimes one uses the notation
\be
\vec{B}_{n+1/2} \equiv \vec{B} \left( \vec{x}_n + \frac{1}{2} \vec{v}_n \Delta t \right)
\ee
to make clear that the magnetic field is evaluated after half a step.

The set of equations listed above allows us to compute the new particle velocity. Note, in the simulations we usually work with the \textit{dimensionless
rigidity vector}
\be
\vec{R} := \frac{1}{\Omega \ell_{\parallel}} \vec{v}
\ee
instead of actual velocities. In the latter formula we have used the parameter $\ell_{\parallel}$ which is a characteristic scale of the turbulence
called the \textit{bendover scale}. It denotes the scale at which we find the turnover from the energy to the inertial range of the spectrum.
Furthermore, in the simulations positions are measured with respect to this scale.

So far we have discussed how the updated velocity can be obtained by employing the Rodrigues formula. However, we also need to know the
updated position. In principle this position is obtained via
\be
\vec{x}_{n+1} = \vec{x}_{n} + \int_{t_n}^{t_n + \Delta t} d t \; \vec{v} \left( t \right).
\label{justbeforemidpoint}
\ee
The velocity vector needed therein is given by
\be
\vec{v} \left( t \right) = e^{\int_{t_n}^t d \tau \; \boldsymbol{M} (\tau)} \vec{v}_n.
\ee
Combining both relation yields
\be
\vec{x}_{n+1} = \vec{x}_{n} + \int_{t_n}^{t_n + \Delta t} d t \; e^{\int_{t_n}^t d \tau \; \boldsymbol{M} (\tau)} \vec{v}_n.
\ee
Therein the matrix $\boldsymbol{M}$ has to be evaluated at the time $\tau$ which is restricted by $t_n \leq \tau \leq t_n + \Delta t$.
As in Eq. (\ref{MinRodrigues}) we used midpoint rule and replace the matrix $\boldsymbol{M}$ by a constant corresponding to the frozen
field approximation. The $\tau$-integral is now straightforward giving us for the position update
\be
\vec{x}_{n+1} = \vec{x}_{n} + \int_{t_n}^{t_n + \Delta t} d t \; e^{\left( t - t_n \right) \boldsymbol{M}} \vec{v}_n
\ee
where the matrix $\boldsymbol{M}$ is given by Eq. (\ref{MinRodrigues}). Using the substitution $\tau = t - t_n$ allows us to write
\be
\vec{x}_{n+1} = \vec{x}_{n} + \int_{0}^{ \Delta t} d \tau \; e^{\boldsymbol{M} \tau} \vec{v}_n.
\label{posupdatewithtau}
\ee
The $\tau$-integral therein can be solved exactly by using again the Rodrigues formula or by using the approach presented in
Appendix \ref{moorepenroseinverse}. To be consistent, however, we simply employ midpoint rule again to approximate Eq. (\ref{posupdatewithtau})
via
\be
\vec{x}_{n+1} = \vec{x}_{n} + \Delta t e^{\boldsymbol{M} \Delta t / 2} \vec{v}_n.
\ee
Alternatively, we can write
\be
\vec{x}_{n+1} = \vec{x}_{n} + \Delta t \vec{v}_{n+1/2}
\label{explicitpositionupdate}
\ee
where we have used
\be
\vec{v}_{n+1/2} = e^{\boldsymbol{M} \Delta t / 2} \vec{v}_n.
\ee
To evaluate the exponential therein we can use again Eq. (\ref{RodriguesFormula}) but replace $\Delta t \rightarrow \Delta t / 2$. Alternatively, one
could just use midpoint rule directly in Eq. (\ref{justbeforemidpoint}) also leading to Eq. (\ref{explicitpositionupdate}). 

\begin{figure}[ht]
\centering
\includegraphics[width=0.45\textwidth]{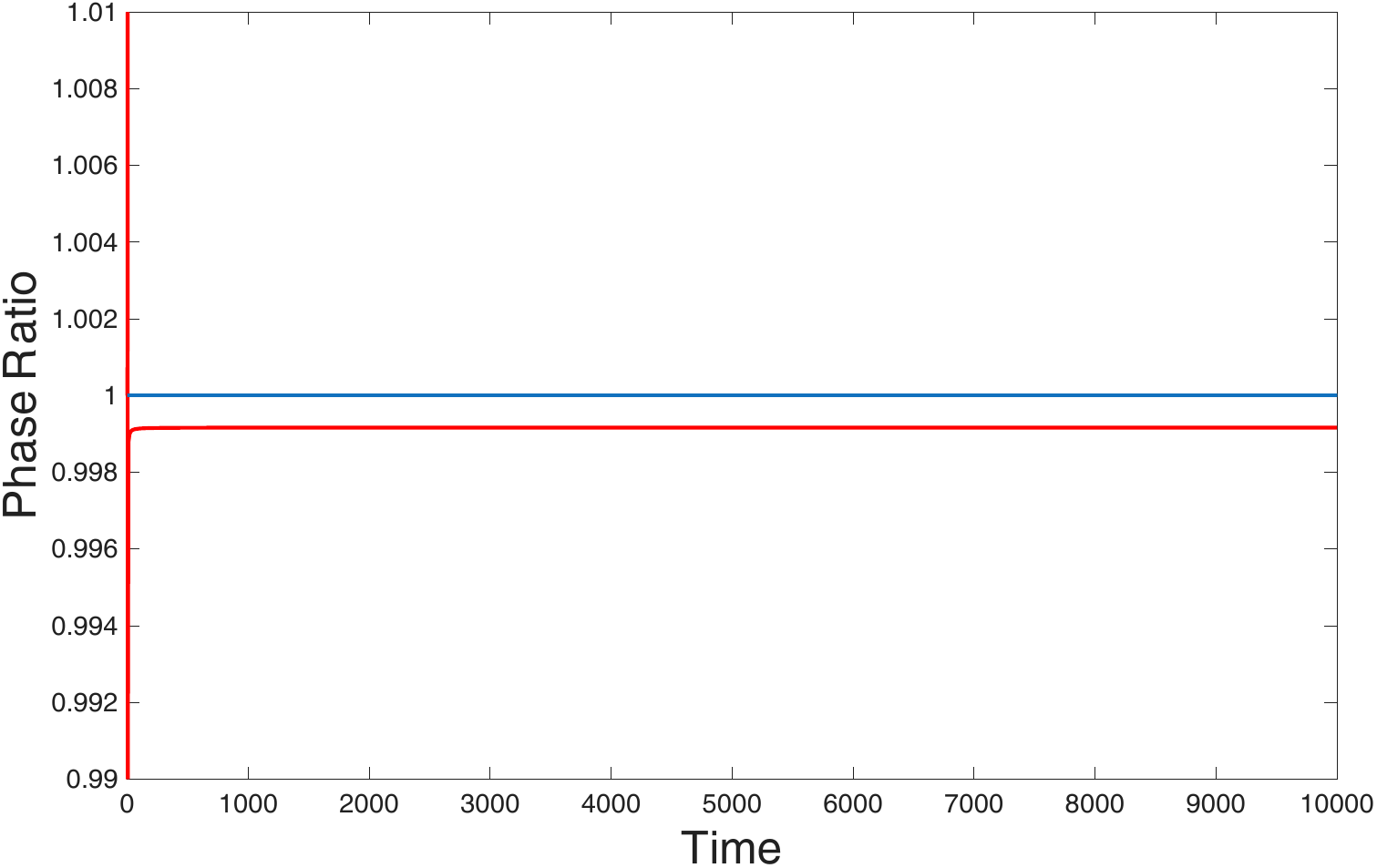}
\caption{The phase defined via Eq. (\ref{definephase}) for the case of a particle interacting with a constant magnetic field pointing in the $z$-direction.
Shown is the ratio of the numerically obtained phase for a step size of $\Delta t = 0.1$ and the exact analytical result for a Rodrigues-based integrator
(blue line) and the Boris scheme (red line).}
\label{ShowPhases}
\end{figure}

Eqs. (\ref{explicitvelocityupdate}) and (\ref{explicitpositionupdate}) can easily be implemented in numerical codes for particle transport. The strength
of this type of approach is that we only made one approximation and that is the one given by Eq. (\ref{MinRodrigues}) corresponding to the frozen field
approximation combined with midpoint rule. Due to the latter approximation, the Rodrigues scheme described above is only correct up to second order.
However, for a constant magnetic field this approximation is not necessary and our computation of the particle orbit is exact. In Fig. \ref{ShowPhases}
we have plotted as an example the ratio of the phase obtained by the Rodrigues scheme and the exact analytical value. This phase is defined and obtained via
\be
\phi := \arctan \left( v_y / v_x \right).
\label{definephase}
\ee
As demonstrated, the Rodrigues integrator provides the exact result for the phase. However, the often mentioned downside of the scheme described above
is that one needs to evaluate the trigonometric functions in the Rodrigues formula at each time step.

\section{The Boris Integrator}

To avoid the occurrence of trigonometric functions in velocity and position updates, a powerful alternative was proposed by Boris (1970).
This integration scheme is widely used in plasma physics and related topics. In fact it is considered to be the \textit{de facto} standard
for particles interacting with electromagnetic fields (see, e.g., Qin et al. (2013)). As starting point we can employ \textit{midpoint rule}
to approximate
\be
\int_{t}^{t + \Delta t} d \tau \; \boldsymbol{M} \left( \tau \right) \approx \Delta t \boldsymbol{M} \left( t + \frac{1}{2} \Delta t \right)
\label{aftermidpoint}
\ee
as it was done above (see Eqs. (\ref{exactvelo2})-(\ref{MinRodrigues})). As in Sect. \ref{RodriguesSect} we, therefore, have employed the frozen
field approximation. To update the velocity, we need to evaluate the matrix exponential in Eq. (\ref{NumMatrixEq}). The first three terms in the
Taylor expansion of this exponential are given by
\be
e^{\boldsymbol{M} \Delta t} \approx \boldsymbol{I} + \boldsymbol{M} \Delta t + \frac{1}{2} \boldsymbol{M}^2 \left( \Delta t \right)^2
\label{Taylorexp}
\ee
which would be correct up to second order in $\Delta t$. However, in the following we use the approximation
\be
e^{\boldsymbol{M} \Delta t} \approx \left( \boldsymbol{I} - \frac{\Delta t}{2} \boldsymbol{M} \right)^{-1} \left( \boldsymbol{I} + \frac{\Delta t}{2} \boldsymbol{M} \right)
\label{Cayleyapprox}
\ee
where we have used the inverse matrix of $\boldsymbol{I} - \boldsymbol{M} \Delta t / 2$. This type of approximation is known as \textit{Cayley approximation}
(see Cayley (1846)), but it is also a special case of a \textit{Pad\'e approximation} (see Pad\'e (1892)). To demonstrate that Eq. (\ref{Cayleyapprox}) is correct
up to second order, we can employ a Taylor expansion by using a \textit{Neumann series} (see Neumann (1877))
\be
\left( \boldsymbol{I} - \boldsymbol{T} \right)^{-1} = \sum_{n=0}^{\infty} \boldsymbol{T}^n
\ee
to get
\bdm
\left( \boldsymbol{I} - \frac{\Delta t}{2} \boldsymbol{M} \right)^{-1} \left( \boldsymbol{I} + \frac{\Delta t}{2} \boldsymbol{M} \right)
& \approx & \left( \boldsymbol{I} + \frac{1}{2} \boldsymbol{M} \Delta t + \frac{1}{4} \boldsymbol{M}^2 \left( \Delta t \right)^2 \right) \left( \boldsymbol{I} + \frac{1}{2} \boldsymbol{M} \Delta t \right) \nonumber\\
& \approx & \boldsymbol{I} + \boldsymbol{M} \Delta t + \frac{1}{2} \boldsymbol{M}^2 \left( \Delta t \right)^2
\edm
in perfect agreement with Eq. (\ref{Taylorexp}). We conclude that up to second order, the Cayley approximation is correct. In Sect. \ref{energyconservation}
we shall demonstrate that Eq. (\ref{Cayleyapprox}) conserves energy whereas Eq. (\ref{Taylorexp}) does not. Within the Boris scheme, we therefore determine
the updated velocity via
\be
\vec{v} \left( t + \Delta t \right) \approx \left( \boldsymbol{I} - \frac{\Delta t}{2} \boldsymbol{M} \right)^{-1} \left( \boldsymbol{I} + \frac{\Delta t}{2} \boldsymbol{M} \right) \vec{v} \left( t \right).
\label{velocitywithCayley}
\ee
One way of evaluating Eq. (\ref{velocitywithCayley}) is to determine the inverse matrix of $\boldsymbol{I} - \boldsymbol{M} \Delta t / 2$. This is done
step-by-step in Appendix \ref{alternativeborisupdate}. Alternatively, we can work with cross products. This approach is outlined in the following. First,
we write Eq. (\ref{velocitywithCayley}) as
\be
\left( \boldsymbol{I} - \frac{\Delta t}{2} \boldsymbol{M} \right) \vec{v} \left( t + \Delta t \right) \approx  \left( \boldsymbol{I} + \frac{\Delta t}{2} \boldsymbol{M} \right) \vec{v} \left( t \right).
\label{noinverseanymore}
\ee
It is convenient to define
\be
\vec{v}^{\;\prime} := \left( \boldsymbol{I} + \frac{\Delta t}{2} \boldsymbol{M} \right) \vec{v} \left( t \right).
\label{definevprime1}
\ee
It follows from the way how we introduced the matrix $\boldsymbol{M}$ (see Eqs. (\ref{NewtonLEq})-(\ref{thematrixMwithdBz})) that
\be
\vec{v}^{\;\prime} = \vec{v} \left( t \right) + \frac{\Omega \Delta t}{2 B_0} \vec{v} \left( t \right) \times \vec{B}.
\label{definevprime2}
\ee
Analogously, we can use this to rewrite the left-hand side of Eq. (\ref{noinverseanymore}) giving us
\be
\vec{v} \left( t + \Delta t \right) - \frac{\Omega \Delta t}{2 B_0} \left[ \vec{v} \left( t + \Delta t \right) \times \vec{B} \right] = \vec{v}^{\;\prime}.
\label{equationwithvprime}
\ee
This corresponds to a vector equation for $\vec{v} \left( t + \Delta t \right)$ that we need to solve. First we multiply Eq. (\ref{equationwithvprime})
by the magnetic field vector $\vec{B}$ via a cross product to get
\be
\vec{v} \left( t + \Delta t \right) \times \vec{B} - \frac{\Omega \Delta t}{2 B_0} \left[ \vec{v} \left( t + \Delta t \right) \times \vec{B} \right] \times \vec{B}
= \vec{v}^{\;\prime} \times \vec{B}.
\ee
By employing the \textit{Grassmann identity}
\be
\vec{a} \times \left( \vec{b} \times \vec{c} \right) = \left( \vec{a} \cdot \vec{c} \right) \vec{b} - \left( \vec{a} \cdot \vec{b} \right) \vec{c}
\label{doublecross}
\ee
we can write this as
\be
\vec{v} \left( t + \Delta t \right) \times \vec{B} + \frac{\Omega \Delta t}{2 B_0} \vec{B}^2 \vec{v} \left( t + \Delta t \right)
- \frac{\Omega \Delta t}{2 B_0} \left[ \vec{v} \left( t + \Delta t \right) \cdot \vec{B} \right] \vec{B} 
= \vec{v}^{\;\prime} \times \vec{B}.
\label{aftergrassmann}
\ee
The velocity component parallel with respect to the magnetic field vector does not change and, thus, we have
\be
\vec{v} \left( t + \Delta t \right) \cdot \vec{B} = \vec{v} \left( t \right) \cdot \vec{B}.
\ee
Using this in Eq. (\ref{aftergrassmann}) yields
\be
\vec{v} \left( t + \Delta t \right) \times \vec{B} + \frac{\Omega \Delta t}{2 B_0} \vec{B^2} \vec{v} \left( t + \Delta t \right)
- \frac{\Omega \Delta t}{2 B_0} \left[ \vec{v} \left( t \right) \cdot \vec{B} \right] \vec{B} 
= \vec{v}^{\;\prime} \times \vec{B}.
\ee
The latter relation can now be used to replace the second term in Eq. (\ref{equationwithvprime}). We derive after some rearranging
\be
\vec{v} \left( t + \Delta t \right) \left[ 1 + \frac{\Omega^2 \Delta t^2}{4 B_0^2} \vec{B}^2 \right]
= \vec{v}^{\;\prime} + \frac{\Omega^2 \Delta t^2}{4 B_0^2} \left[ \vec{v} \left( t \right) \cdot \vec{B} \right] \vec{B}
+ \frac{\Omega \Delta t}{2 B_0} \vec{v}^{\;\prime} \times \vec{B}.
\label{got1plussquare}
\ee
We can rewrite this further. To do this we multiply Eq. (\ref{definevprime2}) by the magnetic field to get the relation
\be
\vec{v}^{\;\prime} \times \vec{B} = \vec{v} \left( t \right) \times \vec{B} + \frac{\Omega \Delta t}{2 B_0} \left[ \vec{v} \left( t \right) \times \vec{B} \right] \times \vec{B}.
\ee
Using Eq. (\ref{doublecross}) once more allows us to write
\be
\frac{\Omega \Delta t}{2 B_0} \left[ \vec{v} \left( t \right) \cdot \vec{B} \right] \vec{B}
= \left[ \vec{v}^{\;\prime} \times \vec{B} - \vec{v} \left( t \right) \times \vec{B} \right] + \frac{\Omega \Delta t}{2 B_0} \vec{B}^2 \vec{v} \left( t \right).
\ee
This can now be used in Eq. (\ref{got1plussquare}) to obtain
\be
\vec{v} \left( t + \Delta t \right) \left[ 1 + \frac{\Omega^2 \Delta t^2}{4 B_0^2} \vec{B}^2 \right]
= \vec{v}^{\;\prime} + \frac{\Omega \Delta t}{2 B_0} \left[ \vec{v}^{\;\prime} \times \vec{B} - \vec{v} \left( t \right) \times \vec{B} \right]
+ \frac{\Omega^2 \Delta t^2}{4 B_0^2} \vec{B}^2 \vec{v} \left( t \right) + \frac{\Omega \Delta t}{2 B_0} \vec{v}^{\;\prime} \times \vec{B}.
\ee
Combining the first term on the right-hand side and the second cross product gives us the velocity vector $\vec{v} \left( t \right)$
which follows from Eq. (\ref{definevprime2}). The first and the third cross products are identical and can, therefore, be combined giving us
\be
\vec{v} \left( t + \Delta t \right) \left[ 1 + \frac{\Omega^2 \Delta t^2}{4 B_0^2} \vec{B}^2 \right]
= \vec{v} \left( t \right) \left[ 1 + \frac{\Omega^2 \Delta t^2}{4 B_0^2} \vec{B}^2 \right] + \frac{\Omega \Delta t}{B_0} \vec{v}^{\;\prime} \times \vec{B}.
\ee
We finally derive for the updated velocity
\be
\vec{v} \left( t + \Delta t \right)
= \vec{v} \left( t \right) + \frac{\Omega \Delta t / B_0}{1 + \omega^2 \Delta t^2 / 4} \vec{v}^{\;\prime} \times \vec{B}
\label{updatedvwithBoris}
\ee
where we have used Eq. (\ref{definelittleomega}) also. Or, we can write the updated velocity via
\be
\vec{v} \left( t + \Delta t \right) = \vec{v} \left( t \right)
+ \frac{\Delta t}{1 + \omega^2 \Delta t^2 / 4} \left( \boldsymbol{M} + \frac{\Delta t}{2} \boldsymbol{M}^2 \right) \vec{v} \left( t \right)
\label{BorisvelostepwithM}
\ee
as demonstrated in Appendix \ref{alternativeborisupdate}. However, we need to keep in mind that the magnetic field has to be evaluated
after half a step (see Eq. (\ref{aftermidpoint})). Therefore, for the use in codes, we write
\be
\vec{v}_{n+1/2} = \vec{v}_{n-1/2} + \frac{\Omega \Delta t / B_0}{1 + \omega^2 \Delta t^2 / 4} \vec{v}^{\;\prime} \times \vec{B}_n
\label{vnplus12}
\ee
with
\be
\vec{v}^{\;\prime} = \vec{v}_{n-1/2} + \frac{\Omega \Delta t}{2 B_0} \vec{v}_{n-1/2} \times \vec{B}_n
\ee
where we have used the notation
\be
\vec{B}_n = \vec{B}_n \left[ \vec{x}_n \right] = \vec{B}_n \left[ \vec{x} \left( t_n \right) \right].
\ee

The updated position is found via
\be
\vec{x} \left( t + \Delta t \right) = \vec{x} \left( t \right) + \int_{t}^{t + \Delta t} d \tau \; \vec{v} \left( \tau \right).
\ee
Using again \textit{midpoint rule} allows us to approximate
\be
\int_{t}^{t + \Delta t} d \tau \; \vec{v} \left( \tau \right) \approx \Delta t \vec{v} \left( t + \frac{1}{2} \Delta t \right)
\ee
so that
\be
\vec{x} \left( t + \Delta t \right) = \vec{x} \left( t \right) + \Delta t \vec{v} \left( t + \frac{1}{2} \Delta t \right).
\ee
Alternatively, this can be written as
\be
\vec{x}_{n+1} = \vec{x}_n + \Delta t \vec{v}_{n+1/2}
\ee
where $\vec{v}_{n+1/2}$ is the updated velocity as provided by Eq. (\ref{vnplus12}).

Compared to the Rodrigues integrator we do not need to evaluate trigonometric functions. However, a Boris integrator does not provide the exact result
for the phase as the Rodrigues scheme does. In Fig. \ref{ShowPhases} we have compared the phases obtained by the two discussed integration methods to
demonstrate this.

Over years several modifications of the Boris scheme have been proposed. There is, for instance, a fourth-order Boris scheme which requires to make
three Boris steps (see Yoshida (1990)). It seems, however, that for the problem described in the current article and for the typically step sizes used
($\Delta t \approx 0.01$) higher-order methods do not provide an improvement for a given run time of the computer. Other possible improvements of the
Boris scheme focus on the inclusion of electric fields (see, for instance, Vay (2008)). However, in the current article we only explore how energetic
particles interact with a purely magnetic system and such possible improvements are not relevant. In cases where one studies stochastic acceleration
due to electric fields (see, e.g., Tautz et al. (2013)) such improvements could become important.

\section{Energy Conservation}\label{energyconservation}

A crucial problem in the simulations described in the current paper is that one needs to consider very long integration times. Therefore, it is important
that our integrators conserve energy. In the following we demonstrate step-by-step that both considered integration methods conserve energy perfectly well.
Both integrators update the velocity via
\be
\vec{v} \left( t + \Delta t \right) = \boldsymbol{R} \vec{v} \left( t \right)
\label{introduceR}
\ee
where the form of the matrix $\boldsymbol{R}$ depends on the used method. To conserve kinetic energy we need
$\left| \vec{v} \left( t + \Delta t \right) \right|^2 = \left| \vec{v} \left( t \right) \right|^2$. By using Eq. (\ref{introduceR})
we can write the left-hand side as
\be
\left| \vec{v} \left( t + \Delta t \right) \right|^2 = \vec{v}^{\mathsf{T}} \left( t + \Delta t \right) \vec{v} \left( t + \Delta t \right)
= \vec{v}^{\mathsf{T}} \left( t \right) \boldsymbol{R}^{\mathsf{T}} \boldsymbol{R} \vec{v} \left( t \right).
\ee
Therefore, energy is conserved when $\boldsymbol{R}^{\mathsf{T}} \boldsymbol{R} = \boldsymbol{I}$ meaning that the matrix $\boldsymbol{R}$ has to be orthogonal.
To explore whether this relation is true for the different approaches discussed in the current article, we need the transpose of the matrix $\boldsymbol{M}$.
It follows directly from Eq. (\ref{thematrixMwithdBz}) that
\be
\boldsymbol{M}^{\mathsf{T}} = - \boldsymbol{M}
\label{Mtranspose}
\ee
meaning the matrix $\boldsymbol{M}$ is \textit{skew-symmetric}. For the case where the original matrix exponential is used (see, e.g., Eq. (\ref{NumMatrixEq}))
we have $\boldsymbol{R} = \exp{(\boldsymbol{M} \Delta t)}$. One can easily show with the help of Eq. (\ref{Mtranspose}) that
\be
\boldsymbol{R}^{\mathsf{T}} \boldsymbol{R} = e^{\boldsymbol{M}^{\mathsf{T}} \Delta t} e^{\boldsymbol{M} \Delta t}
= e^{\left( \boldsymbol{M}^{\mathsf{T}} + \boldsymbol{M} \right) \Delta t} = \boldsymbol{I}.
\ee
Since the Rodrigues formula is an exact representation of the matrix exponential, a Rodrigues integrator conserves energy perfectly well.

For a Boris integrator, which corresponds to rotations, we expect the same. In the following we present a step-by-step proof. It follows
from Eq. (\ref{BorisvelostepwithM}) that
\be
\boldsymbol{R} = \boldsymbol{I} + \frac{\Delta t}{1 + \omega^2 \Delta t^2 / 4} \left( \boldsymbol{M} + \frac{\Delta t}{2} \boldsymbol{M}^2 \right).
\ee
Using Eq. (\ref{Mtranspose}) yields
\be
\boldsymbol{R}^{\mathsf{T}} = \boldsymbol{I} + \frac{\Delta t}{1 + \omega^2 \Delta t^2 / 4} \left( - \boldsymbol{M} + \frac{\Delta t}{2} \boldsymbol{M}^2 \right).
\ee
From this we can derive
\be
\boldsymbol{R}^{\mathsf{T}} \boldsymbol{R}
= \boldsymbol{I} + \frac{\Delta t^2}{1 + \omega^2 \Delta t^2 / 4} \boldsymbol{M}^2
- \frac{\Delta t^2}{\left( 1 + \omega^2 \Delta t^2 / 4 \right)^2} \left( \boldsymbol{M}^2 + \frac{1}{4} \Delta t^2 \boldsymbol{M}^4 \right).
\ee
Employing the second line of Eq. (\ref{Mrelations}) and performing some straightforward algebra yields indeed $\boldsymbol{R}^{\mathsf{T}} \boldsymbol{R} = \boldsymbol{I}$.
Therefore, the Boris integrator, although based on an approximation of the matrix exponential, conserves energy perfectly well. Note, if we would use the
Taylor expansion given by Eq. (\ref{Taylorexp}) instead of Eq. (\ref{Cayleyapprox}), energy would not be conserved. This is the reason why we have employed
the Cayley approximation. It conserves energy exactly, but is still correct up to second order. This is an important feature of the Boris scheme.

\section{Performing the Simulations}

To compare the performance of the different integrators (Rodrigues versus Boris), we consider energetic particle transport in slab turbulence. In this case
the magnetic field has the form
\be
\vec{B} = \delta B_x \left( z \right) \hat{x} + \delta B_y \left( z \right) \hat{y} + B_0 \hat{z}
\ee
where $B_0$ is the mean magnetic field pointing in the $z$-direction. The turbulent magnetic field in slab turbulence depends only on the coordinate along the
mean field and is perfectly transverse due to the solenoidal constraint. The slab model used here as an example is supported by solar wind data indicating that
solar wind turbulence consists of slab modes and so-called two-dimensional modes (see Matthaeus et al. (1990)).

In the simulations one needs to create the magnetic field. There are two ways of doing this:
\begin{enumerate}
\item In grid-based methods magnetic field vectors are created everywhere in space based on a superposition of modes. In the equation of motion, ones uses interpolation
to determine the magnetic field at the particle position. Such grid-based methods have been widely used in the past to explore how energetic particles interact
with magnetic turbulence (see, e.g., Qin et al. (2002a), Qin et al. (2002b), and Pommois et al. (2007)).
\item One computes the magnetic field only at the position where the particle currently resides. However, in this case the magnetic field has to be
computed anew at each time step. This method is described in detail in Giacalone \& Jokipii (1999) as well as Tautz (2010).
\end{enumerate}
Both methods have advantages and disadvantages (see Heusen (2017) for some discussions). In the current article we employ the second method which is sometimes
called \textit{continuous approach} (see again Heusen (2017)).

For the slab model one creates the magnetic field via the formula
\be
\delta \vec{B} \left( z \right) = \sqrt{2} \delta B \sum_{n=0}^{N-1} A \left( k_n \right) \hat{\xi}_n \cos \left( k_n z + \beta_n \right).
\label{createB}
\ee
Therein we have used the total turbulent magnetic field $\delta B$ as well as the function $A \left( k_n \right)$ which is described below. Furthermore,
we have used the polarization vector $\hat{\xi}_n=\left( -\sin \phi_n, \cos \phi_n, 0 \right)$ where the set of angles $\phi_n$ is created via random numbers.
Eq. (\ref{createB}) also contains the phases $\beta_n$ which are created via random numbers as well. The modes $k_n$ are obtained via
\be
k_n = k_{min} \left[ \frac{k_{max}}{k_{min}} \right]^{n/(N-1)} \quad \textnormal{with} \quad n=0,1,\dots,N-1
\label{Expkn}
\ee
corresponding to a logarithmic spacing. The latter formula contains the smallest and largest modes $k_{min}$ and $k_{max}$, respectively. The parameter $N$
denotes the number of modes used in the simulations.

In Eq. (\ref{createB}) we have used the function $A \left( k_n \right)$ which is defined via
\be
A^2 \big( k_n \big) = G \big( k_n \big) \Delta k_n \left( \sum_{m=0}^{N-1} G \big( k_m \big) \Delta k_m \right)^{-1}.
\label{defineA2}
\ee
The latter function satisfies
\be
\sum_{n=0}^{N-1} A^2 \big( k_n \big) = 1
\label{normforA}
\ee
corresponding to normalization. For the spectrum of the slab modes we employ the form
\be
G \left( k_n \right) = \left( 1 + k_n^2 \right)^{-s/2}.
\label{Bieberspectrum}
\ee
This spectrum describes a smooth turnover from the energy range, where the spectrum is assumed to be constant, to the inertial range. In the latter part
of the spectrum we have $\propto k_n^{-s}$ where $s$ is the inertial range spectral index. The spectrum used here was originally proposed by Bieber et al. (1994).
The formulas listed above allow us to create the turbulent magnetic field at an arbitrary position in space $z$.

Note, combining Eqs. (\ref{Expkn}) and (\ref{Bieberspectrum}) allows one to create a specific spectrum ranging from the smallest wave number $k_{min}$
to the largest wave number $k_{max}$ using a logarithmic spacing. This can easily be changed. If one desires using a spectrum obtained from observations,
for instance, one has two options:
\begin{enumerate}
\item Keep Eq. (\ref{Expkn}) to create the grid points in wave number space and replace the spectrum given by Eq. (\ref{Bieberspectrum}) by something which fits the data.
\item Use the data points from the observations directly. Then one would no longer work with a logarithmic spacing and Eq. (\ref{Expkn}) is not needed anymore.
\end{enumerate}
The second option can be problematic because the logarithmic spacing is useful numerically because it allows one the cover the whole wave number space equally.
But it is still possible to work like this. For the remained of the paper, we create the spectrum as described above.

To run the simulations we have considered $10^5$ particles, $256$ wave modes, and two values of the magnetic rigidity, namely $R=0.1$ and $R=10$. The smallest
mode considered is $k_{min} = 0.001$ and the largest mode is $k_{max} = 1000$. This choice guarantees that our spectrum has proper energy and inertial ranges.
For the inertial range spectral index we have used $s = 5/3$ as originally proposed by Kolmogorov (1941). For the total turbulent magnetic field we have set
$\delta B^2 = 0.20 B_0^2$ as suggested by Bieber et al. (1994) and Bieber et al. (1996).

The magnetic field, created as described above, can now be used in the numerical integrators discussed in the previous sections. After obtaining the particle position
at a given time, we can determine the running diffusion coefficients in the two directions of space. Usually a diffusion coefficient is defined as time-derivative of
the mean square displacement of the orbits. However, this would lead to very noisy data and, thus, we compute
\be
d_{\parallel} \left( t \right) = \frac{1}{2 t} \langle \left( \Delta z \right)^2 \quad\quad\textnormal{and}\quad\quad d_{\perp} \left( t \right) = \frac{1}{2 t} \langle \left( \Delta x \right)^2
\ee
instead.

\begin{figure}[ht]
\centering
\includegraphics[width=0.45\textwidth]{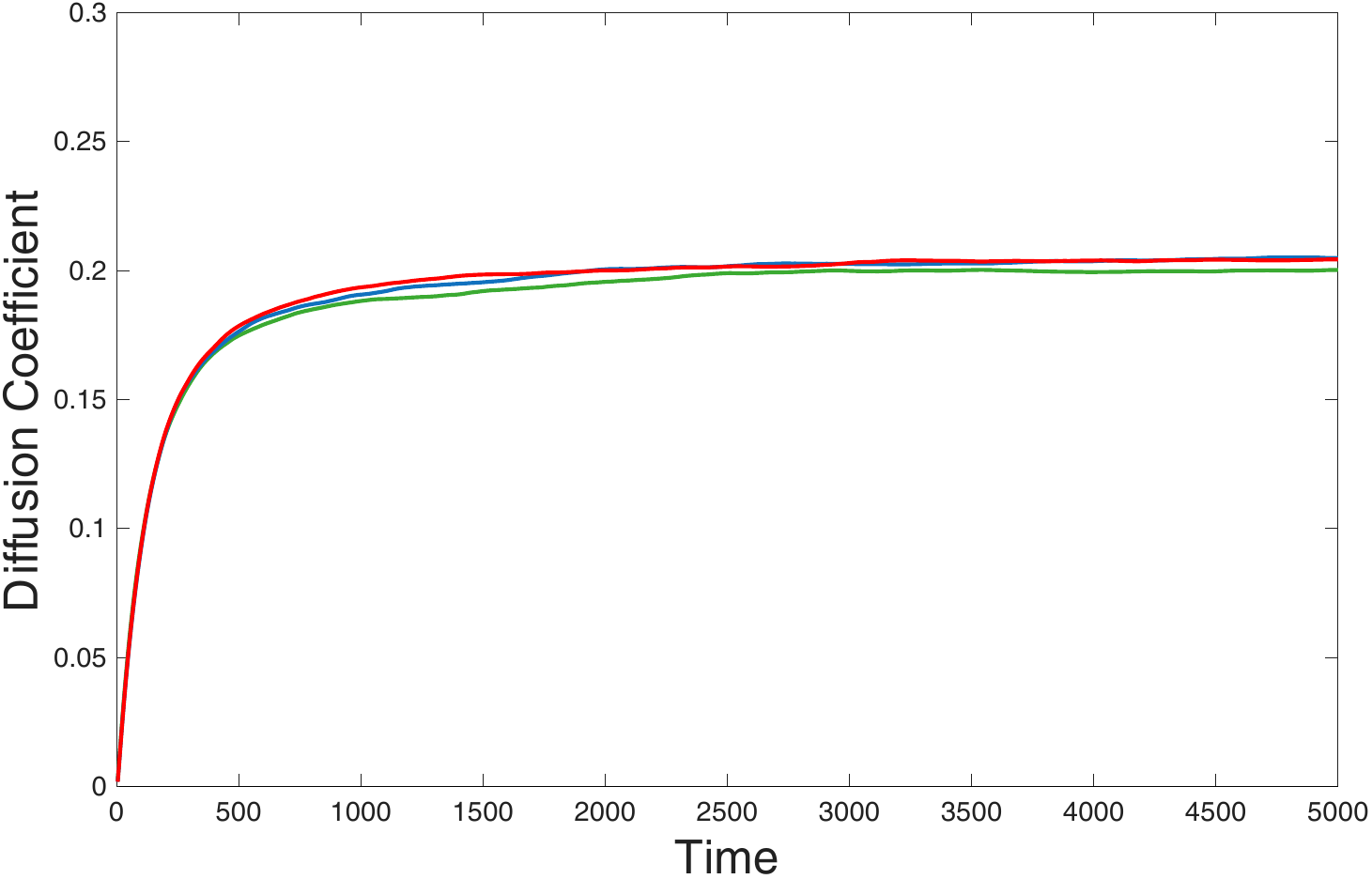}
\caption{The parallel diffusion coefficient obtained from the simulations by employing the Rodrigues integrator. We have performed the simulations
for different values of the step size, namely $\Delta t = 1$ (green line), $\Delta t = 0.1$ (blue line), and $\Delta t = 0.01$ (red line). Here we
have considered the case of small rigidities $R=0.1$.}
\label{ParallelRodriguesR0p1}
\end{figure}

\begin{figure}[ht]
\centering
\includegraphics[width=0.45\textwidth]{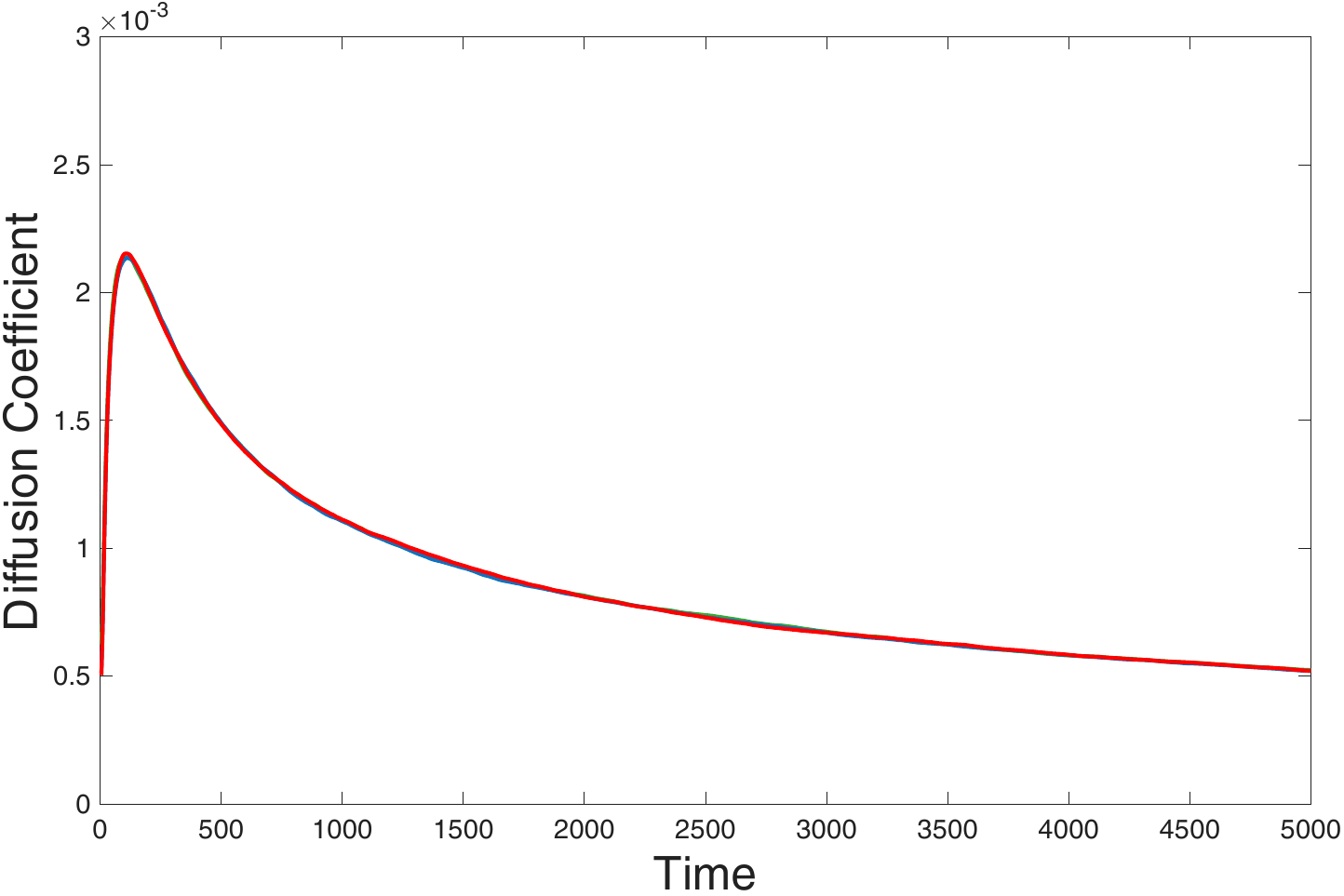}
\caption{Caption is as in Fig. \ref{ParallelBorisR0p1} but we have considered perpendicular diffusion. Note, all three curves are in coincidence.}
\label{PerpendicularRodriguesR0p1}
\end{figure}

\begin{figure}[ht]
\centering
\includegraphics[width=0.45\textwidth]{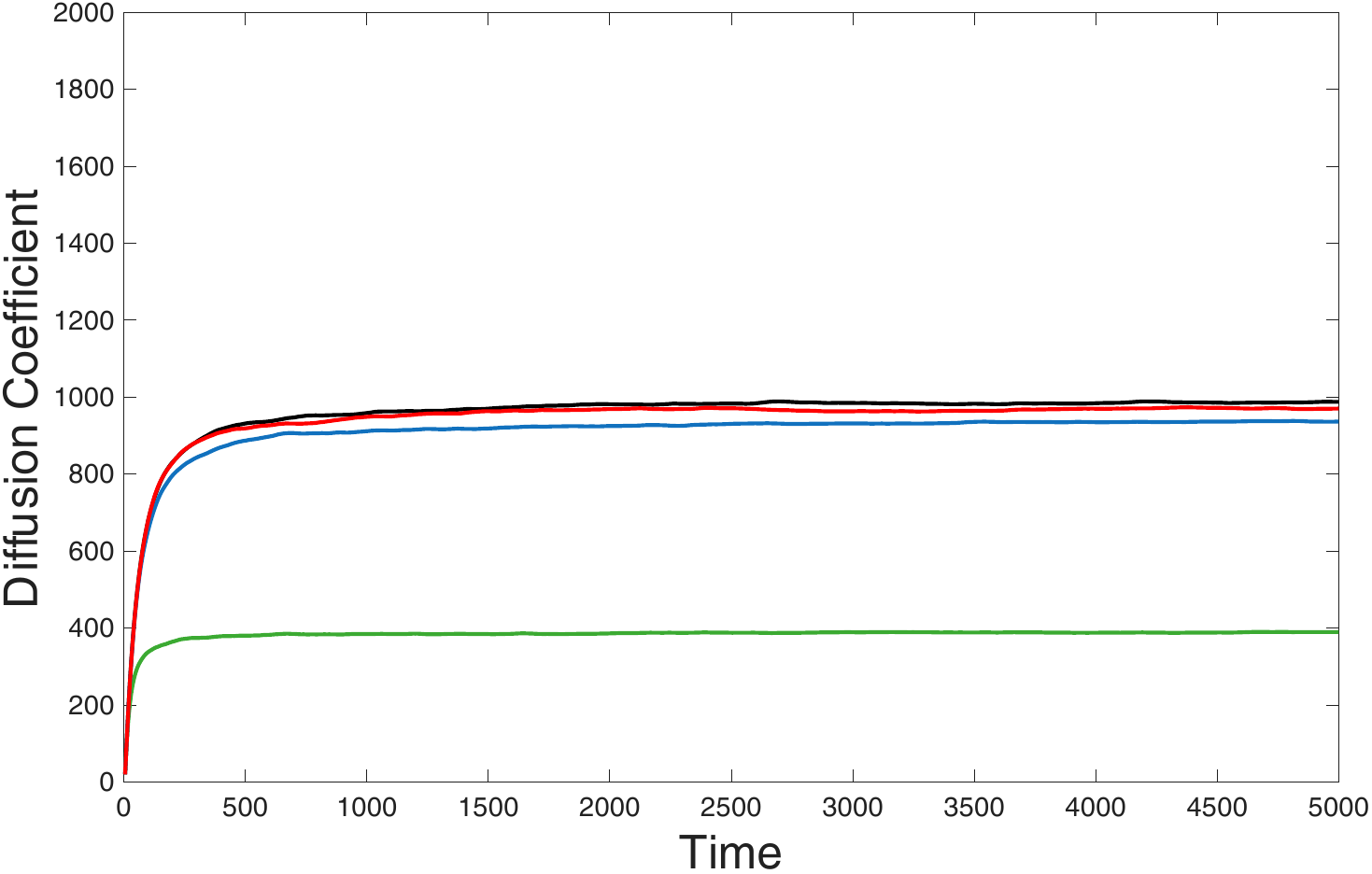}
\caption{The parallel diffusion coefficient obtained from the simulations by employing the Rodrigues integrator. We have performed the simulations
for different values of the step size, namely $\Delta t = 1$ (green line), $\Delta t = 0.1$ (blue line), $\Delta t = 0.01$ (red line), and
$\Delta t = 0.001$ (black line). Here we have considered the case of large rigidities $R=10$.}
\label{ParallelRodriguesR10}
\end{figure}

\begin{figure}[ht]
\centering
\includegraphics[width=0.45\textwidth]{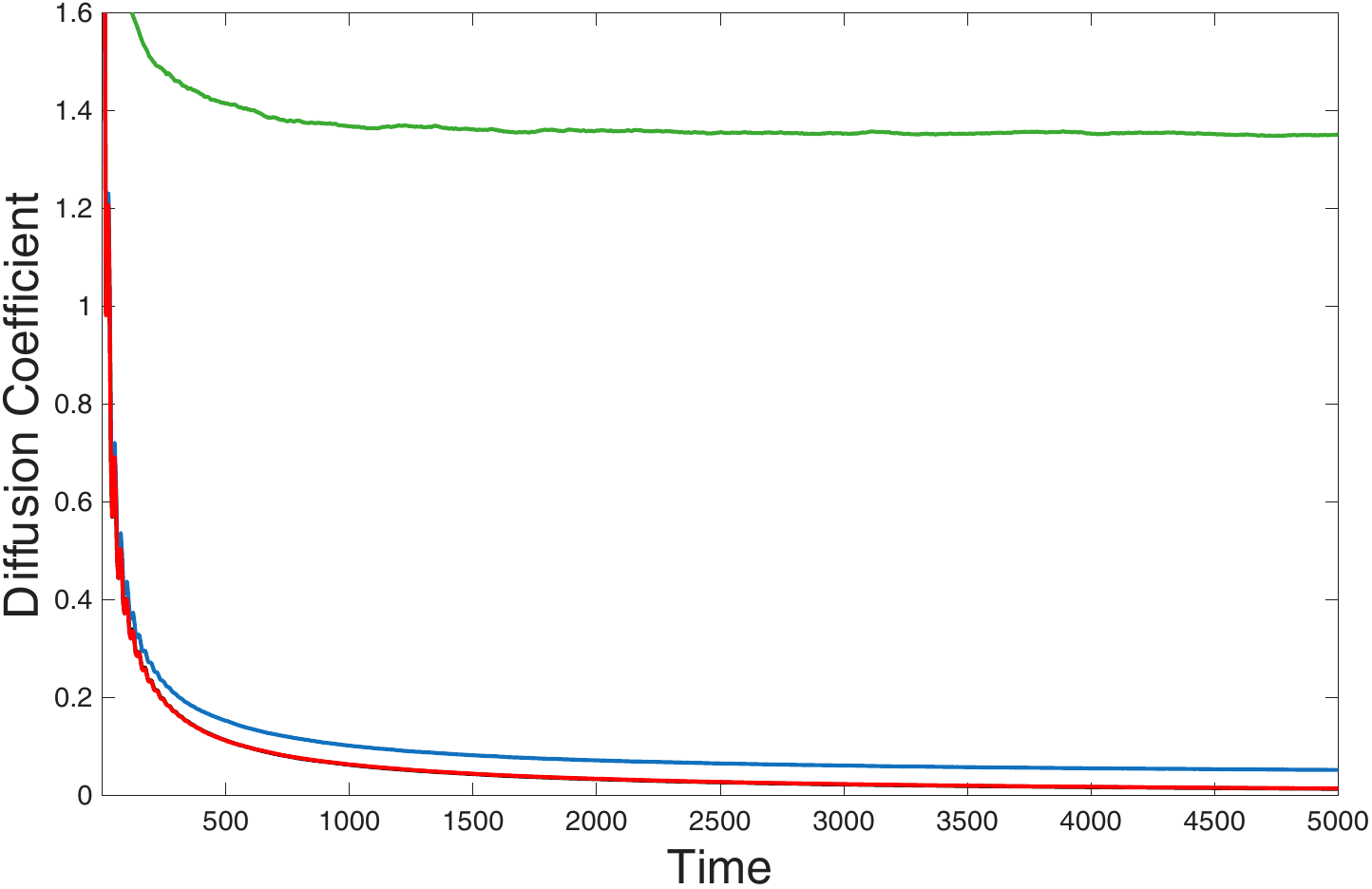}
\caption{Caption is as in Fig. \ref{ParallelBorisR10} but we have considered perpendicular diffusion. Note, red and black lines are in coincidence.}
\label{PerpendicularRodriguesR10}
\end{figure}

\begin{figure}[ht]
\centering
\includegraphics[width=0.45\textwidth]{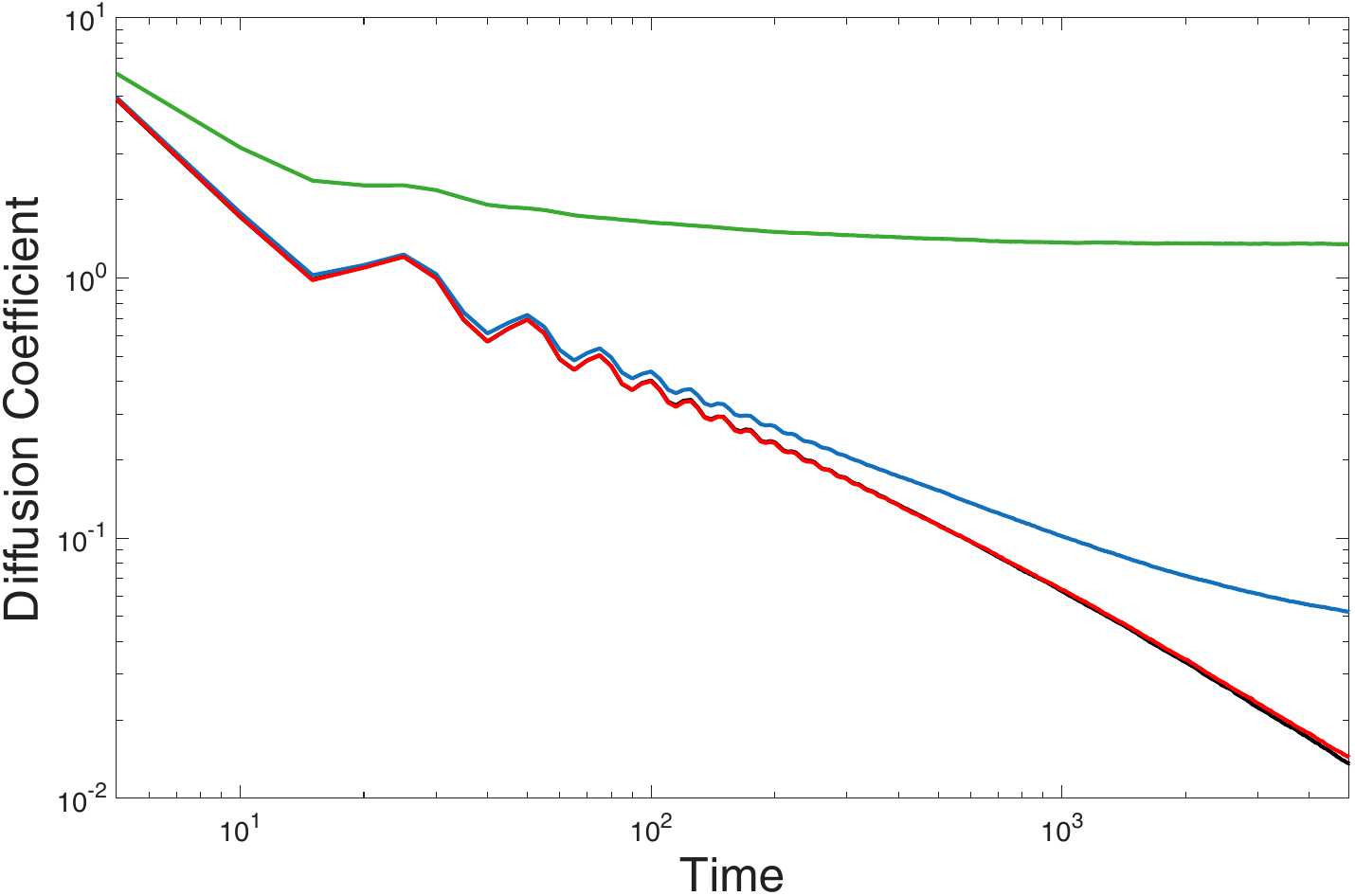}
\caption{Caption is as in Fig. \ref{PerpendicularRodriguesR10} but results are shown as double-logarithmic plot.}
\label{PerpendicularRodriguesR10log}
\end{figure}

\begin{figure}[ht]
\centering
\includegraphics[width=0.45\textwidth]{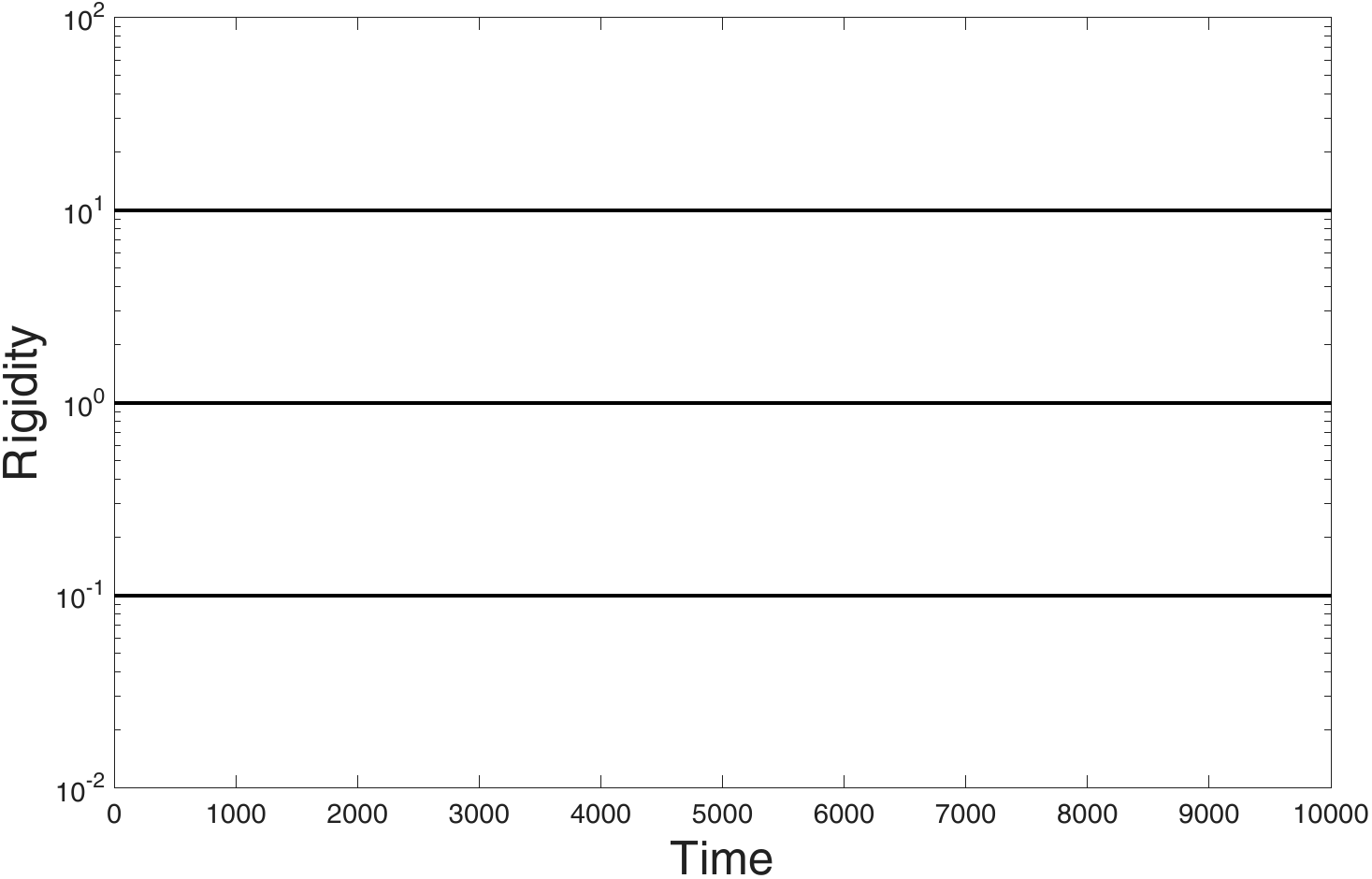}
\caption{Rigidity conservation for a Rodrigues integrator.}
\label{RigidityRodrigues}
\end{figure}

To determine the accuracy of the two integration schemes, we performed the simulations by considering different sizes of the time steps $\Delta t$.
Of course, the result is more accurate for smaller values of $\Delta t$. In Figs. \ref{ParallelRodriguesR0p1}-\ref{PerpendicularRodriguesR10log} we
have shown parallel and perpendicular diffusion coefficients obtained by employing the Rodrigues integrator. Simulations in pure slab turbulence have
been performed before and the results for the diffusion coefficients are well-known (see Qin et al. (2002a)). For parallel transport we find normal
diffusion after the initial ballistic regime. Perpendicular transport is initially ballistic as well but after overcoming this regime perpendicular
transport becomes sub-diffusive with $\langle \left( \Delta x \right)^2 \rangle \propto t^{1/2}$ as described very well by a process called
\textit{compound sub-diffusion} (see, e.g., Webb et al. (2006) and Shalchi (2019)). The purpose of the plots shown here is to explore the efficiency
of the used numerical integration scheme. One can clearly see that the result converges quickly for a step size much smaller than one. A Rodrigues
integrator with $\Delta t = 0.1$ should provide a very accurate result as long as the rigidity $R$ is smaller than one. However, for the high-rigidity
case, where $R=10$, obtaining an accurate result requires a smaller step size (e.g., $\Delta t = 0.01$). Fig. \ref{RigidityRodrigues} shows the time
evolution of the rigidity of individual particles having different initial rigidities, namely $R=0.1$, $R=1$, and $R=10$. As demonstrated, energy
is conserved perfectly well as predicted in Sect. \ref{energyconservation}.


\begin{figure}[ht]
\centering
\includegraphics[width=0.45\textwidth]{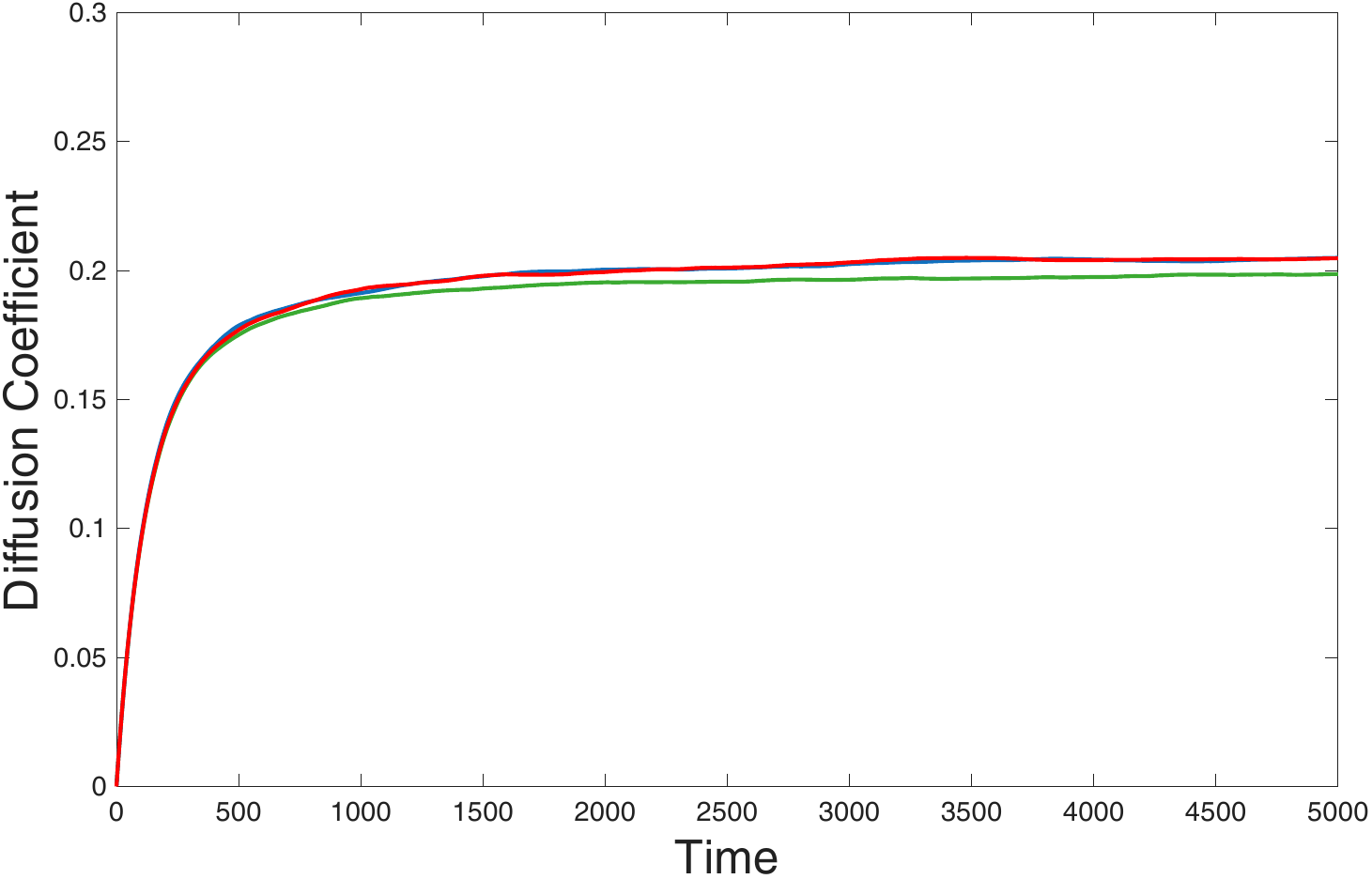}
\caption{The parallel diffusion coefficient obtained from the simulations by employing the Boris integrator. We have performed the simulations for different
values of the step size, namely $\Delta t = 1$ (green line), $\Delta t = 0.1$ (blue line), and $\Delta t = 0.01$ (red line). Here we have considered the case
of small rigidities $R=0.1$. Note, red and blue lines are in perfect coincidence.}
\label{ParallelBorisR0p1}
\end{figure}

\begin{figure}[ht]
\centering
\includegraphics[width=0.45\textwidth]{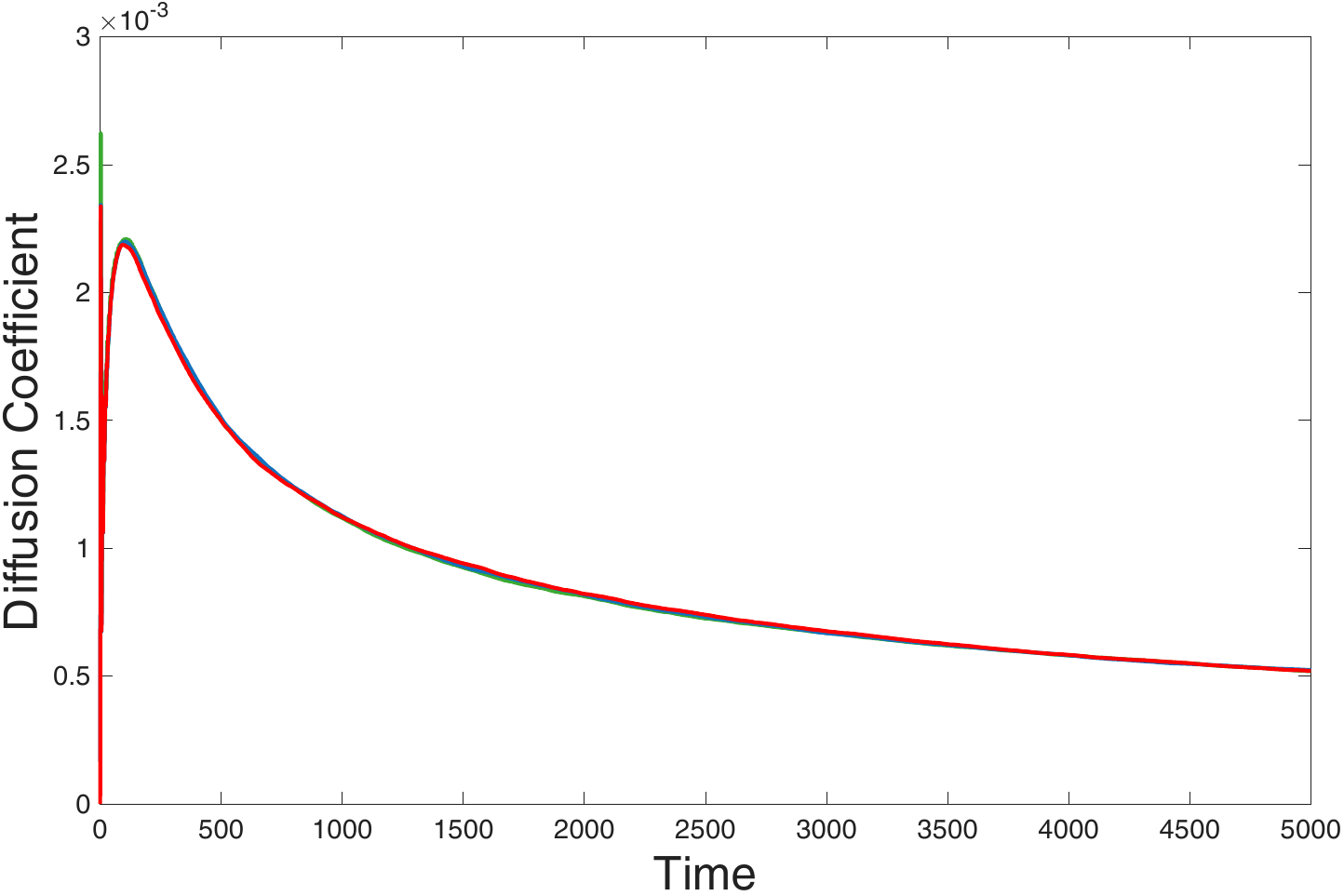}
\caption{Caption is as in Fig. \ref{ParallelBorisR0p1} but we have considered perpendicular diffusion. Note, green, red, and blue lines are in perfect coincidence.}
\label{PerpendicularBorisR0p1}
\end{figure}

\begin{figure}[ht]
\centering
\includegraphics[width=0.45\textwidth]{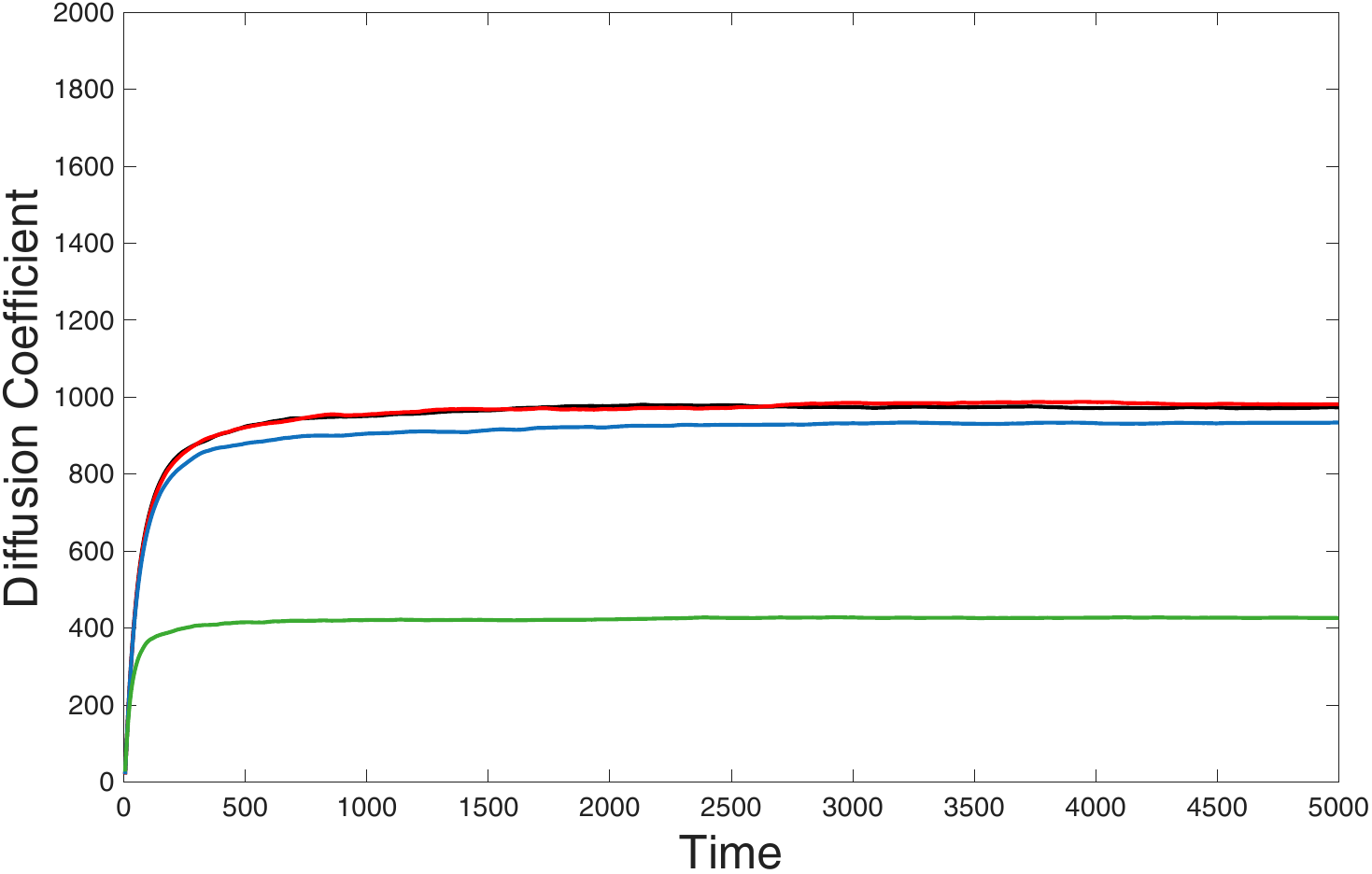}
\caption{The parallel diffusion coefficient obtained from the simulations by employing the Boris integrator. We have performed the simulations for different
values of the step size, namely $\Delta t = 1$ (green line), $\Delta t = 0.1$ (blue line), $\Delta t = 0.01$ (red line), and $\Delta t = 0.001$ (black line).
Here we have considered the case of large rigidities $R=10$.}
\label{ParallelBorisR10}
\end{figure}

\begin{figure}[ht]
\centering
\includegraphics[width=0.45\textwidth]{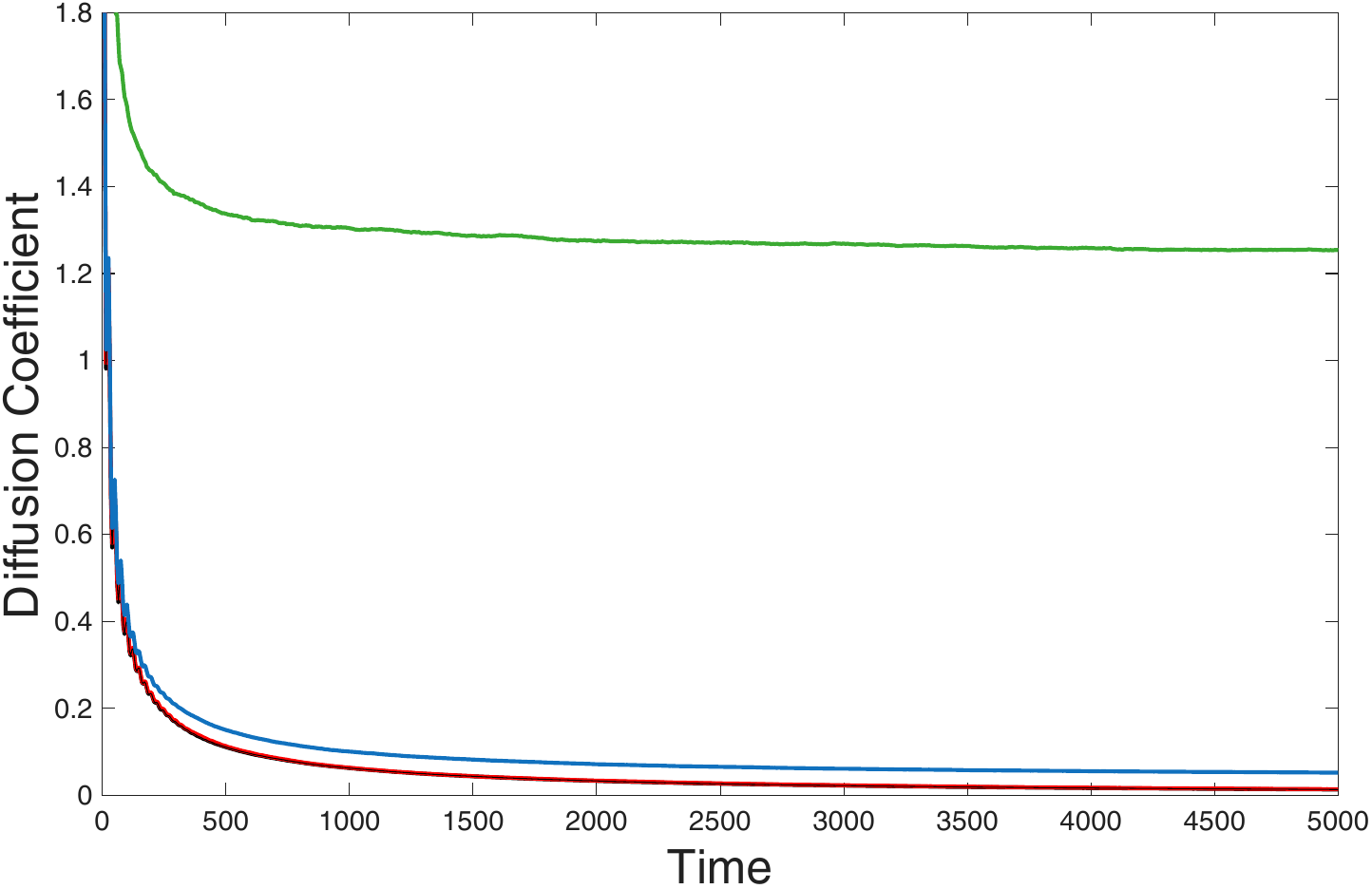}
\caption{Caption is as in Fig. \ref{ParallelBorisR10} but we have considered perpendicular diffusion.}
\label{PerpendicularBorisR10}
\end{figure}

\begin{figure}[ht]
\centering
\includegraphics[width=0.45\textwidth]{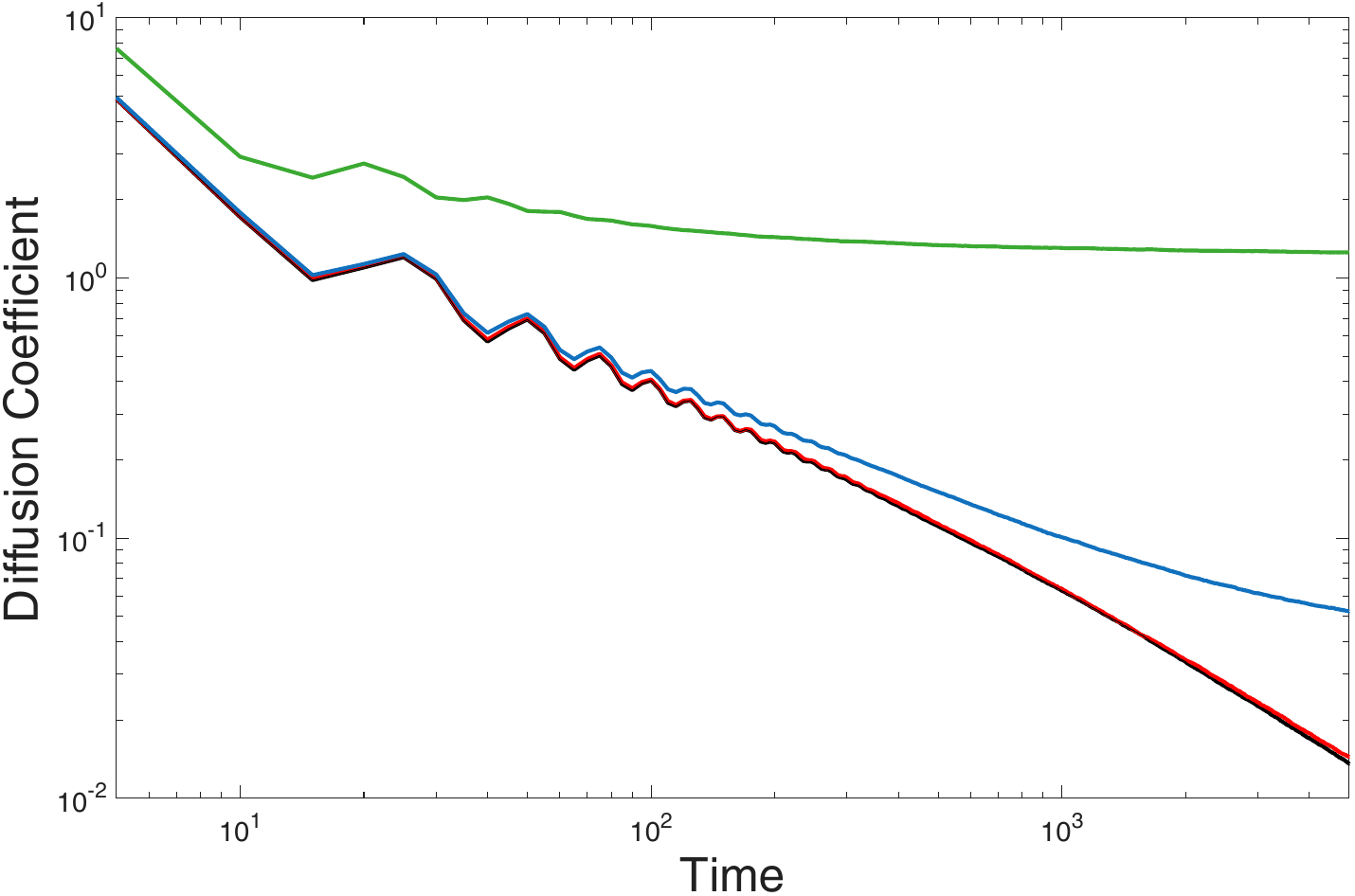}
\caption{Caption is as in Fig. \ref{PerpendicularBorisR10} but results are shown as double-logarithmic plot.}
\label{PerpendicularBorisR10log}
\end{figure}

\begin{figure}[ht]
\centering
\includegraphics[width=0.45\textwidth]{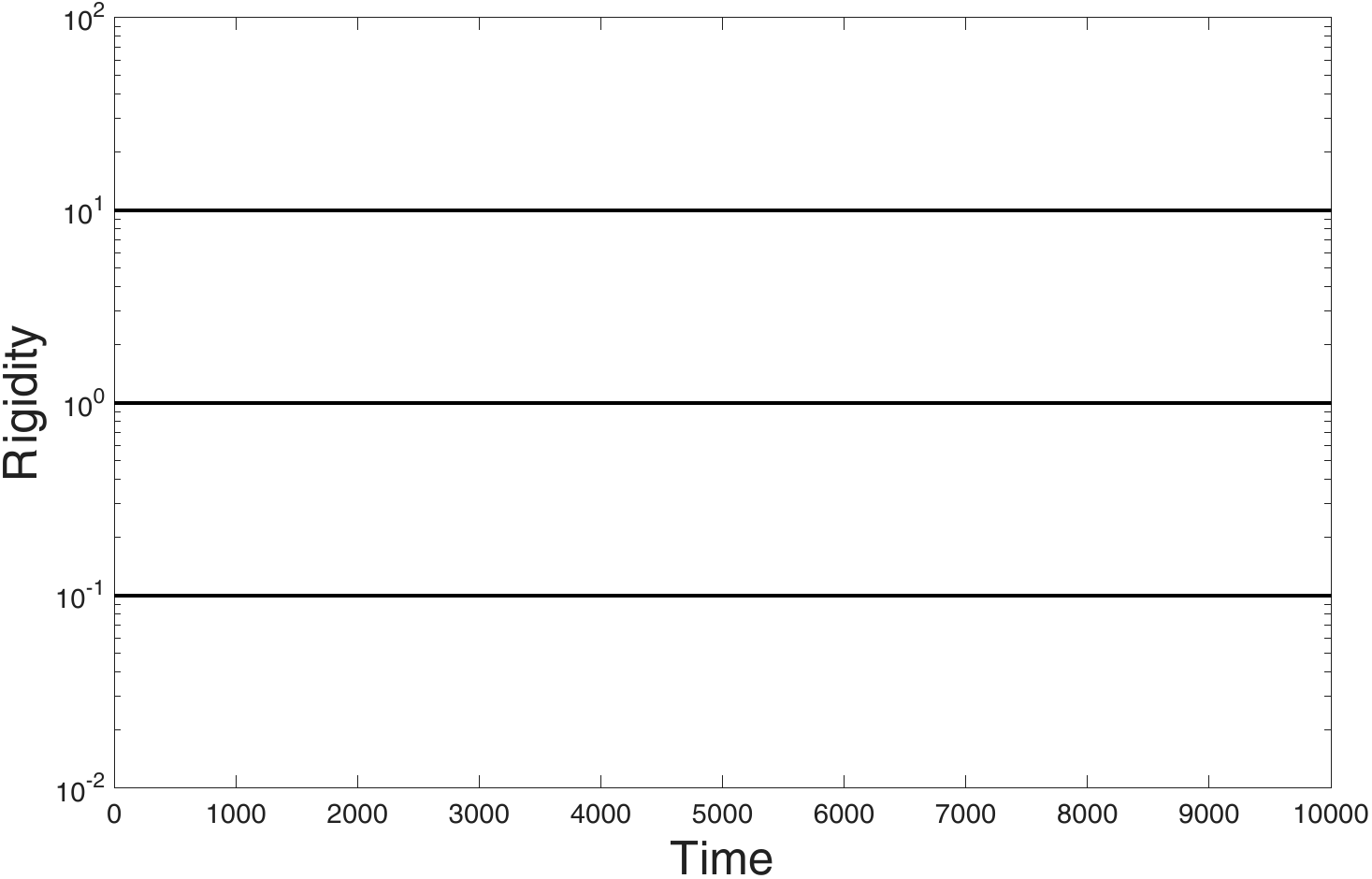}
\caption{Rigidity conservation for a Boris integrator.}
\label{RigidityBoris}
\end{figure}

Figs. \ref{ParallelBorisR0p1}-\ref{PerpendicularBorisR10log} show parallel and perpendicular diffusion coefficients for the same parameter values, but we have
now used a Boris integrator. Fig. \ref{RigidityBoris} demonstrates energy conservation for the Boris integration scheme. The conclusions we can draw from those
plots are similar compared to the one for a Rodrigues integrator. For all considered cases both schemes provide similar results, converge to the final result with
a similar pace, and conserve energy perfectly well. In theory the Rodrigues scheme should be more accurate because the matrix exponential is evaluated by using
the exact Rodrigues formula instead of the Cayley approximation. The run times of both schemes are also very similar. The fact that the Rodrigues formula contains
trigonometric functions has no noticeable impact on run times.

%
%

%\newpage

%
%
%
\section{Summary and Conclusion}

The current article compares two integration schemes one can use to solve the Newton-Lorentz equation for a particle interacting with magnetic turbulence
numerically. Both integrators can be derived by rewriting the Newton-Lorentz equation as matrix equation (see Eq. (\ref{diffeq1})) which allows one to
express the solution for the new particle velocity by a matrix exponential (see Eq. (\ref{exactvelo})). However, the matrix in the exponential is a
time-integrated matrix containing the magnetic field components. This cannot be evaluated exactly for a time-dependent field and one has to rely on
approximations. First we have employed the frozen field approximation meaning that the magnetic field is assumed to be constant while we perform a single
time step. This is done by employing midpoint rule (see Eqs. (\ref{MinRodrigues}) and (\ref{aftermidpoint})) leading to second-order accuracy of the schemes.
However, even if the frozen field approximation is used, one still has to deal with a matrix exponential. In the current paper we have discussed two options
to deal with this problem:
\begin{enumerate}
\item We can employ the Rodrigues formula (\ref{RodriguesFormula}) which is an exact relation. The disadvantage of this approach is that this formula contains
trigonometric functions which have to be evaluated numerically during each time step. This can be numerically expensive in certain scenarios.
\item We can use the Cayley approximation (\ref{Cayleyapprox}) which is not exact but leads to a formula for the updated velocity which does not contain any
trigonometric function. The corresponding integration scheme is known as Boris integrator. Although it is based on an additional approximation, it conserves
energy perfectly well.
\end{enumerate}
It is usually thought that the second method is much more efficient because, while the first method is in theory more accurate, the Boris scheme is much faster.
However, if one explores a particle interacting with a turbulent field which is purely magnetic, this statement is no longer true. The reason is that at every
step one computes the magnetic field via Eq. (\ref{createB}). This evaluation of the magnetic field is computationally much more expensive than computing two
trigonometric functions. Therefore, in theory, a Rodrigues integrator should be slightly better compared to a Boris integrator. For the instance, the former
scheme provides an exact result for the phase in the case where the particle interacts with a constant magnetic field (see Fig. \ref{ShowPhases} of the current
article).

Of course one has to keep in mind that this superiority of the Rodrigues scheme emerges if one uses the continues approach to create the turbulent magnetic field.
In grid-based methods, for instance, the fact that the Rodrigues integrator contains trigonometric functions could nullify this advantage. Overall both approaches,
the integration scheme based on the Rodrigues formula and Boris' method, are very efficient if one desires to describe energetic particles interacting with magnetic
turbulence numerically.

\vspace*{0.2cm}

\begin{acknowledgments}
\textit{Support by the Natural Sciences and Engineering Research Council (NSERC) of Canada is acknowledged.}
\end{acknowledgments}

\appendix
\section{The Moore-Penrose Inverse}\label{moorepenroseinverse}

For the Rodrigues approach the updated position can also be obtained by using
\be
\int_{0}^{\Delta t} d \tau \; e^{\boldsymbol{M} \tau} = \boldsymbol{M}^{+} \left( e^{\boldsymbol{M} \Delta t} - \boldsymbol{I} \right) + \Delta t \boldsymbol{P}_0
\label{introMoorePenrose}
\ee
where we have used the \textit{Moore-Penrose pseudoinverse} (see Moore (1920) and Penrose (1955)) which satisfies
\be
\boldsymbol{M}^{+} = - \frac{B_0^2}{\Omega^2 \left( B_x^2 + B_y^2 + B_z^2 \right)} \boldsymbol{M}
\equiv - \frac{1}{\omega^2} \boldsymbol{M}.
\label{MoorePenrose}
\ee
Note, this pseudoinverse is defined so that
\be
\boldsymbol{M} \boldsymbol{M}^{+} = \boldsymbol{I} - \hat{B} \otimes \hat{B}.
\ee
Furthermore, we have
\be
\boldsymbol{M} \boldsymbol{M}^{+} = \boldsymbol{M}^{+} \boldsymbol{M} = - \frac{1}{\omega^2} \boldsymbol{M}^2
\label{MMplus}
\ee
which follows directly from Eq. (\ref{MoorePenrose}). By combining Eqs. (\ref{Mrelations}) and (\ref{MoorePenrose}) we can easily derive
\be
\boldsymbol{M}^{+} \boldsymbol{M}^2 = - \frac{1}{\omega^2} \boldsymbol{M}^3 = \boldsymbol{M}.
\label{MplusM2}
\ee
To determine the matrix $\boldsymbol{P}_0$ in Eq. (\ref{introMoorePenrose}), we consider the derivative
\be
\frac{d}{d \Delta t} \left[ \boldsymbol{M}^{+} \left( e^{\boldsymbol{M} \Delta t} - \boldsymbol{I} \right) + \Delta t \boldsymbol{P}_0 \right] 
= \boldsymbol{M}^{+} \boldsymbol{M} e^{\boldsymbol{M} \Delta t} + \boldsymbol{P}_0.
\ee
Therein we can use Eq. (\ref{MMplus}) and the Rodrigues formula (\ref{RodriguesFormula}) to get
\be
\frac{d}{d \Delta t} \left[ \boldsymbol{M}^{+} \left( e^{\boldsymbol{M} \Delta t} - \boldsymbol{I} \right) + \Delta t \boldsymbol{P}_0 \right]
= - \frac{1}{\omega^2} \boldsymbol{M}^2 \left[ \boldsymbol{I} + \frac{\sin \left( \omega \Delta t \right)}{\omega} \boldsymbol{M}
+ \frac{1 - \cos \left( \omega \Delta t \right)}{\omega^2} \boldsymbol{M}^2 \right] + \boldsymbol{P}_0.
\ee
Employing Eq. (\ref{Mrelations}) yields
\be
\frac{d}{d \Delta t} \left[ \boldsymbol{M}^{+} \left( e^{\boldsymbol{M} \Delta t} - \boldsymbol{I} \right) + \Delta t \boldsymbol{P}_0 \right] 
= - \frac{1}{\omega^2} \boldsymbol{M}^2 + \frac{\sin \left( \omega \Delta t \right)}{\omega} \boldsymbol{M} + \frac{1 - \cos \left( \omega \Delta t \right)}{\omega^2} \boldsymbol{M}^2 + \boldsymbol{P}_0.
\ee
Using the Rodrigues formula once more allows us to write
\be
\frac{d}{d \Delta t} \left[ \boldsymbol{M}^{+} \left( e^{\boldsymbol{M} \Delta t} - \boldsymbol{I} \right) + \Delta t \boldsymbol{P}_0 \right] 
= e^{\boldsymbol{M} \Delta t} - \boldsymbol{I} - \frac{1}{\omega^2} \boldsymbol{M}^2 + \boldsymbol{P}_0.
\ee
Because this has to be equal to the exponential, we can read off
\be
\boldsymbol{P}_0 = \boldsymbol{I} + \frac{1}{\omega^2} \boldsymbol{M}^2.
\ee
The matrix $\boldsymbol{P}_0$ is called the \textit{projector onto the null space} of $\boldsymbol{M}$ (see again Penrose (1955)). We can finally write
\be
\int_{0}^{\Delta t} d \tau \; e^{\boldsymbol{M} \tau}
= \boldsymbol{M}^{+} \left( e^{\boldsymbol{M} \Delta t} - \boldsymbol{I} \right) + \Delta t \left( \boldsymbol{I} + \frac{1}{\omega^2} \boldsymbol{M}^2 \right).
\label{excactexpintegral}
\ee
This formula allows one to compute a time-integral over an exponential containing the product of time and a constant matrix $\boldsymbol{M}$.

\section{Alternative Derivation of the Boris Velocity Update}\label{alternativeborisupdate}

According to Eq. (\ref{velocitywithCayley}) we have
\be
\vec{v} \left( t + \Delta t \right) \approx \left( \boldsymbol{I} - \frac{1}{2} \Delta t \boldsymbol{M} \right)^{-1} \left( \boldsymbol{I} + \frac{1}{2} \Delta t \boldsymbol{M} \right) \vec{v} \left( t \right).
\ee
The complication therein is that one needs to know the inverse matrix of $\boldsymbol{I} - \boldsymbol{M} \Delta t / 2$ where $\boldsymbol{M}$ is given by
Eq. (\ref{thematrixMwithdBz}). First, we need to determine the determinant of this matrix which is found to be
\be
\det \left[ \boldsymbol{I} - \frac{1}{2} \Delta t \boldsymbol{M} \right] = 1 + \omega^2 \Delta t^2 / 4
\ee
where $\omega$ is given by Eq. (\ref{definelittleomega}).

To determine the inverse, we also need to compute the corresponding \textit{adjunct matrix}. However, the result should be some function of the
matrix $\boldsymbol{M}$. Due to Eq. (\ref{Mrelations}) we can make the \textit{ansatz}
\be
\left( \boldsymbol{I} - \frac{1}{2} \Delta t \boldsymbol{M} \right)^{-1}
= \frac{1}{1 + \omega^2 \Delta t^2 / 4} \left[ \alpha \boldsymbol{I} + \beta \boldsymbol{M} + \gamma \boldsymbol{M}^2 \right].
\label{introduceabc}
\ee
To determine the three coefficients $\alpha$, $\beta$, and $\gamma$ therein, we use Eq. (\ref{introduceabc}) in
\be
\left( \boldsymbol{I} - \frac{\Delta t}{2} \boldsymbol{M} \right) \left( \boldsymbol{I} - \frac{\Delta t}{2} \boldsymbol{M} \right)^{-1} = \boldsymbol{I}.
\ee
After some straightforward algebra we get $\alpha = 1 + \omega^2 \Delta t^2 / 4$, $\beta = \Delta t / 2$, and $\gamma = \Delta t^2 / 4$.
Therewith, we find for the inverse
\be
\left( \boldsymbol{I} - \frac{\Delta t}{2} \boldsymbol{M} \right)^{-1} = \boldsymbol{I} + \frac{1}{1 + \omega^2 \Delta t^2 / 4} \frac{\Delta t}{2} \boldsymbol{M}
+ \frac{1}{1 + \omega^2 \Delta t^2 / 4} \frac{\Delta t^2}{4} \boldsymbol{M}^2.
\ee
With the derived relation we can now determine the needed matrix product
\be
\left( \boldsymbol{I} - \frac{\Delta t}{2} \boldsymbol{M} \right)^{-1} \left( \boldsymbol{I} + \frac{\Delta t}{2} \boldsymbol{M} \right)
= \left[ \boldsymbol{I} + \frac{1}{1 + \omega^2 \Delta t^2 / 4} \frac{\Delta t}{2} \boldsymbol{M}
+ \frac{1}{1 + \omega^2 \Delta t^2 / 4} \frac{\Delta t^2}{4} \boldsymbol{M}^2 \right] \left( \boldsymbol{I} + \frac{\Delta t}{2} \boldsymbol{M} \right).
\ee
% \bdm
% & & \left( \boldsymbol{I} - \frac{\Delta t}{2} \boldsymbol{M} \right)^{-1} \left( \boldsymbol{I} + \frac{\Delta t}{2} \boldsymbol{M} \right) \nonumber\\
% & = & \left[ \boldsymbol{I} + \frac{1}{1 + \omega^2 \Delta t^2 / 4} \frac{\Delta t}{2} \boldsymbol{M}
% + \frac{1}{1 + \omega^2 \Delta t^2 / 4} \frac{\Delta t^2}{4} \boldsymbol{M}^2 \right] \left( \boldsymbol{I} + \frac{\Delta t}{2} \boldsymbol{M} \right).
% \edm
Again we can employ the first line of Eq. (\ref{Mrelations}) to work this out. We derive
\be
\left( \boldsymbol{I} - \frac{\Delta t}{2} \boldsymbol{M} \right)^{-1} \left( \boldsymbol{I} + \frac{\Delta t}{2} \boldsymbol{M} \right)
= \boldsymbol{I} + \frac{\Delta t}{1 + \omega^2 \Delta t^2 / 4} \boldsymbol{M} + \frac{\Delta t^2 / 2}{1 + \omega^2 \Delta t^2 / 4} \boldsymbol{M}^2.
\ee
Therewith the updated velocity can be written as
\be
\vec{v} \left( t + \Delta t \right) = \vec{v} \left( t \right)
+ \frac{\Delta t}{1 + \omega^2 \Delta t^2 / 4} \boldsymbol{M} \left( \boldsymbol{I} + \frac{\Delta t}{2} \boldsymbol{M} \right) \vec{v} \left( t \right).
\label{updatedwithfullmatrix}
\ee
Comparing this with Eq. (\ref{definevprime1}) allows us to write
\be
\vec{v} \left( t + \Delta t \right) = \vec{v} \left( t \right)
+ \frac{\Delta t}{1 + \omega^2 \Delta t^2 / 4} \boldsymbol{M} \vec{v}^{\;\prime}.
\ee
When the matrix $\boldsymbol{M}$ acts on a vector, it generates a cross product and a factor $\Omega$. Therefore, we find
\be
\vec{v} \left( t + \Delta t \right) = \vec{v} \left( t \right)
+ \frac{\Omega \Delta t}{1 + \omega^2 \Delta t^2 / 4} \vec{v}^{\;\prime} \times \vec{B}
\ee
in perfect agreement with Eq. (\ref{updatedvwithBoris}).

{}


\begin{thebibliography}{}

\bibitem[Bieber et al.(1994)]{bieber94}
Bieber, J.W., Matthaeus, W.H., Smith, C.W., Wanner, W., Kallenrode, M.-B., Wibberenz, G.: Astrophys. J. \textbf{420}, 294 (1994)

\bibitem[Bieber et al.(1996)]{bieber96}
Bieber, J.W., Wanner, W., Matthaeus, W.H.: J. Geophys. Res. \textbf{101}, 2511 (1996)
%Dominant two-dimensional solar wind turbulence with implications for cosmic ray transport,

\bibitem[Boris(1970)]{boris70}
Boris, J.: Relativistic plasma simulation-optimization of a hybrid code, Proceedings of the Fourth Conference on Numerical
Simulation of Plasmas. Naval Research Laboratory, Washington, D.C., 3-67 (1970)

\bibitem[Burden \& Faires(2011)]{burdenfaires11}
Burden, R.L., Faires, J.D.: Numerical Analysis, 9th ed., Brooks/Cole, Cengage Learning, Boston (2011)

\bibitem[Cayley(1846)]{cayley1846}
Cayley, A.: Journal f\"ur die reine und angewandte Mathematik \textbf{32} 119 (1846)

\bibitem[Dosch \& Shalchi(2010)]{doschshal2010}
Dosch, A., Shalchi A.: Adv. Space Res. \textbf{46} 1208 (2010)

\bibitem[Effenberger et al.(2025)]{Effenberger2025}
Effenberger, F., Walter, D., Fichtner, H., et al.:
%Open Issues in Non-Gaussian Transport and Acceleration of Charged Energetic Particles in Space and Astrophysical Plasmas.
Space Sci. Rev. \textbf{221}, 75 (2025)

\bibitem[Els \& Engelbrecht(2024)]{ElsEng24}
Els, P., Engelbrecht, N.E.: Astrophys. J. \textbf{969}, 1 (2024)
%On Calculating Diffusion Coefficients Numerically in Synthetic Turbulence Using Particle Pushers, Astrophys. J. Supplement \textbf{969}, 1 (2024)

\bibitem[Engelbrecht \& Wolmarans(2020)]{EngWol20}
Engelbrecht, E.N., Wolmarans, C.P.: Adv. Space Res. \textbf{66}, 2722 (2020) 

\bibitem[Engelbrecht \& Moloto(2021)]{EngMol21}
Engelbrecht, E.N., Moloto, K.D.: Astrophys. J. \textbf{908}, 167 (2021) 

\bibitem[Ferrand et al.(2014)]{ferretal14}
Ferrand, G., Danos, R., Shalchi, A., Safi-Harb, S., Mendygral, P.: Astrophys. J. \textbf{792}, 133 (2014) 

\bibitem[Fraternale et al.(2022)]{Fraternale22}
Fraternale, F., Adhikari, L., Fichtner, H., Kim, T.K., Kleimann, J., Oughton, S., Pogorelov, N.V., Roytershteyn, V., Smith, C.W., Usmanov, A.V., Zank, G.P., Zhao, L.L.:
Space Sci. Rev. \textbf{218} 50 (2022)

\bibitem[Giacalone \& Jokipii(1999)]{giacjok99}
Giacalone, J., Jokipii, J. R.: Astrophys. J. \textbf{520}, 204 (1999) 
%THE TRANSPORT OF COSMIC RAYS ACROSS A TURBULENT MAGNETIC FIELD

\bibitem[Hairer et al.(2006)]{Haireretal2006}
Hairer, E., Lubich, C., Wanner, G.: Geometric numerical integration: Structure-preserving algorithms for ordinary differential equations (2nd ed.),
Springer (2006)

\bibitem[Heusen \& Shalchi(2016)]{heusha16}
Heusen, M., Shalchi, A.: Astrophys. Space Sci. \textbf{361}, 308 (2016)
%Simulations of energetic particles interacting with nonlinear anisotropic dynamical turbulence, 

\bibitem[Heusen(2017)]{heusen2017}
Heusen, M.H.: Analytical and numerical investigation of energetic particles interacting with turbulent magnetic fields,
PhD Thesis, University of Manitoba (2017)

\bibitem[Hu et al.(2017)]{hu17}
Hu, J., Li, G., Ao, X., Zank, G.P., Verkhoglyadova, O.: J. Geophys. Res. \textbf{122}, 938 (2017) 

\bibitem[Hussein et al.(2015)]{Hussein15}
Hussein, M., Tautz, R.C., Shalchi, A.: J. Geophys. Res. \textbf{120}, 4095 (2015) 
%The influence of different turbulence models on the diffusion coefficients of energetic particles, J. Geophys. Res. \textbf{120}, 4095 (2015)

\bibitem[Hussein \& Shalchi(2016)]{husha16}
Hussein, M., Shalchi, A.: Astrophys. J. \textbf{817}, 136 (2016) 
%Simulations of Energetic Particles Interacting with Dynamical Magnetic Turbulence, Astrophys. J. \textbf{817}, 136 (2016)

\bibitem[Ivascenko et al.(2016)]{Ivas16}
Ivascenko, A., Lange, S., Spanier, F., Vainio, R.: Astrophys. J. \textbf{833}, 223 (2016) 
%Determining pitch-angle diffusion coefficients form test particle simulations, Astrophys. J. \textbf{833}, 223 (2016)

\bibitem[Kolmogorov(1941)]{kol41}
Kolmogorov, A.N.: Dokl. Akad. Nauk. SSSR \textbf{30}, 301 (1941)

\bibitem[Li et al.(2003)]{li03}
Li, G., Zank, G.P., Rice, W.K.M.: J. Geophys. Res. \textbf{108}, 1082 (2003)

\bibitem[Li et al.(2005)]{li05}
Li, G., Zank, G.P., Rice, W.K.M.: J. Geophys. Res. \textbf{110}, A06104 (2005) 

\bibitem[Li et al.(2012)]{lietal12}
Li, G., Shalchi, A., Ao, X., Zank, G.P., Verkhoglyadova, O.P.: Adv. Space Res. \textbf{49}, 1067 (2012) 

\bibitem[Marsden \& Ratiu(1999)]{marsrat99}
Marsden, J.E., Ratiu, T.S.: Introduction to Mechanics and Symmetry. A Basic Exposition of Classical Mechanical Systems, Second edition,
Texts in Applied Mathematics 17, Springer-Verlag, New York, (1999)

\bibitem[Matthaeus et al.(1990)]{matt90}
Matthaeus, W.H., Goldstein, M.L., Aaron, R.D.: J. Geophys. Res. \textbf{95}, 20673 (1990)

\bibitem[Matthaeus et al.(2003)]{matt03}
Matthaeus, W.H., Qin, G., Bieber, J.W., Zank, G.P.: Astrophys. J. \textbf{590}, L53 (2003) 

\bibitem[Moloto \& Engelbrecht(2020)]{moloto20}
Moloto, K.D., Engelbrecht, N.E.: Astrophys. J. \textbf{894}, 121 (2020) 

\bibitem[Moore(1920)]{moore1920}
Moore, E.H.: Bulletin of the American Mathematical Society \textbf{26}, 394 (1920)
%On the reciprocal of the general algebraic matrix, 

\bibitem[Neumann(1877)]{Neumann1877}
Neumann, C.: Untersuchungen \"uber das logarithmische und Newton'sche Potential, Teubner, Leipzig (1877)

\bibitem[Ngobeni et al.(2022)]{Ngobeni22}
Ngobeni, M.D., Potgieter, M.S., Aslam, O.P.M., Bisschoff, D., Ramokgaba, I.I., Ndiitwani, D.C.: Adv. Space Res. \textbf{69}, 2330 (2022) 

\bibitem[Pad\'e(1892)]{pade1892}
Pad\'e, H.: Sur la r\'epresentation approch\'ee d'une fonction par des fractions rationelles (Thesis), Ann. \'Ecole Nor. 3, 3-93 (1892)

\bibitem[Penrose(1955)]{penrose1955}
Penrose, R.: A generalized inverse for matrices, Proceedings of the Cambridge Philosophical Society, 51, 406 (1955)

\bibitem[Perri \& Zimbardo(2007)]{perr07}
Perri, S., Zimbardo, G.: Astrophys. J. \textbf{671}, L177 (2007) 

\bibitem[Perri \& Zimbardo(2009a)]{Perri2009a}
Perri, S., Zimbardo, G.: Astrophys. J. \textbf{693}, L118 (2009a) 

\bibitem[Perri \& Zimbardo(2009b)]{Perri2009b}
Perri, S., Zimbardo G.: Adv. Space Res. \textbf{44}, 465 (2009b) 

\bibitem[Pommois et al.(2007)]{Pomm2007}
Pommois, P., Zimbardo, G., Veltri, P.: Phys. Plasma \textbf{14}, 012311 (2007)

\bibitem[Qin et al.(2002a)]{qin2002a}
Qin, G., Matthaeus, W.H., Bieber, J.W.: Geophys. Res. Lett. \textbf{29}, 1048 (2002a)

\bibitem[Qin et al.(2002b)]{qin02b}
Qin, G., Matthaeus, W.H., Bieber, J.W.: Astrophys. J. \textbf{578}, L117 (2002b)

\bibitem[Qin et al.(2006)]{qinetal06}
Qin, G., Matthaeus, W.H., Bieber, J.W.: Astrophys. J. \textbf{640}, L103 (2006)
%Parallel Diffusion of Charged Particles in Strong Two-dimensional Turbulence, Astrophys. J. \textbf{640}, L103 (2006)

\bibitem[Qin et al.(2018)]{qin18}
Qin, G., Kong, F.-J., Zhang, L.-H.: Astrophys. J. \textbf{860}, 3 (2018) 

\bibitem[Qin et al.(2013)]{Qinetal2013}
Qin, H., Zhang, S., Xiao, J., Liu, J., Sun, Y., Tang, W.M.: Phys. Plasmas \textbf{20}, 084503 (2013)

\bibitem[Reichherzer et al.(2022)]{reich22}
Reichherzer, P., Becker Tjus, J., Zweibel, E. G., et al.: Mon. Not. R. Astron. Soc. \textbf{514}, 2658 (2022)
%Anisotropic cosmic ray diffusion in isotropic Kolmogorov turbulence, MNRAS \textbf{514}, 2658-2666 (2022)

\bibitem[Ripperda et al.(2018)]{Ripperdaetal2018}
Ripperda, B., Bacchini, F., Teunissen, J., Xia, C., Porth, O., Sironi, L., Lapenta, G., Keppens, R.:
Astrophys. J. \textbf{235}, 21 (2018)
%A Comprehensive Comparison of Relativistic Particle Integrators, 

\bibitem[Rodrigues(1840)]{Rodrigues1840}
Rodrigues, O.: Journal de Math\'ematiques Pures et Appliqu\'ees \textbf{5}, 380 (1840)
%Des lois g\'eom\'etriques qui r\'egissent les d\'eplacements d'un syst\'eme solide dans l'espace, et de la variation des coordonn\'ees
%provenant de ces d\'eplacements consid\'er\'es ind\'ependants des causes qui peuvent les produire

\bibitem[Shen \& Qin(2018)]{shenqin18}
Shen, Z.-N., Qin, G.: Astrophys. J. \textbf{854}, 137 (2018) 

\bibitem[Shen et al.(2021)]{Shen21}
Shen, Z., Qin, G., Zuo, P., Wei, F., Xu, X.: Astrophys. J. \textbf{256}, 18 (2021) 

\bibitem[Shalchi(2009)]{shal09book}
Shalchi, A.: Nonlinear Cosmic Ray Diffusion Theories, Astrophysics and Space Science Library, vol. 362, Springer, Berlin (2009)

\bibitem[Shalchi(2010)]{shalchi10unlt}
Shalchi, A.: Astrophys. J. \textbf{720}, L127 (2010) 

\bibitem[Shalchi(2019)]{2019letter}
Shalchi, A.: Astrophys. J. \textbf{881}, L27 (2019)
%Heuristic Description of Perpendicular Diffusion of Energetic Particles in Astrophysical Plasmas.

\bibitem[Shalchi(2020)]{shal2020rev}
Shalchi, A.: Space Sci. Rev. \textbf{216}, 23 (2020)
%Perpendicular Transport of Energetic Particles in Magnetic Turbulence. Space Science Reviews

\bibitem[Shalchi(2021)]{shal2021FLPD}
Shalchi, A.: Astrophys. J. \textbf{923}, 209 (2021) 

\bibitem[Snodin et al.(2022)]{Snodin22}
Snodin, A.P., Jitsuk, T., Ruffolo, D., Matthaeus, W.H.: Astrophys. J. \textbf{932}, 127 (2022) 
%Energetic Particle Perpendicular Diffusion: Simulations and Theory in Noisy Reduced Magnetohydrodynamic Turbulence, Astrophys. J. \textbf{932}, 127 (2022)

\bibitem[Tautz(2010)]{tautz10}
Tautz, R.C.: Comput. Phys. Commun. \textbf{181}, 71 (2010)
%A new simulation code for particle diffusion in anisotropic, large-scale and turbulent magnetic fields, 

\bibitem[Tautz et al.(2013)]{Tautz13}
Tautz, R.C., Lerche, I., Kruse, F.: Astron. Astrophys. \textbf{555}, A101 (2013)

\bibitem[Vay(2008)]{vay08}
Vay, J.-L.: Phys. Plasmas \textbf{15}, 056701 (2008)
%Simulation of beams or plasmas crossing at relativistic velocity

\bibitem[Webb et al.(2006)]{Webbetal06}
Webb, G.M., Zank, G.P., Kaghashvili, E.Kh., le Roux, J.A.: Astrophys. J. \textbf{651}, 211 (2006)

\bibitem[Yoshida(1990)]{yosh90}
Yoshida, H.: Phys. Lett. A \textbf{150}, 262 (1990)
%Construction of higher order symplectic integrators

\bibitem[Zank et al.(2000)]{zank00}
Zank, G.P., Rice, W.K.M., Wu, C.C.: J. Geophys. Res. \textbf{105}, 25079 (2000) 

\bibitem[Zank et al.(2006)]{zanketal06}
Zank, G.P., Li, G., Florinski, V., Hu, Q., Lario, D., Smith, C.W.: J. Geophys. Res. \textbf{111}, A06108 (2006) 

\bibitem[Zank et al.(2019)]{Zanketal2019}
Zank G.P., Nakanotani M., Webb, G.M.: Astrophys. J. \textbf{88}, L116 (2019) 

\bibitem[Zhao et al.(2020)]{Zhaoetal2020}
Zhao, L.L, Zank G.P., Burlaga, L.F.: Astrophys. J. \textbf{900}, 166 (2020) 

\bibitem[Zhao et al.(2024)]{Zhaoetal2024}
Zhao, L.L, Zank G.P., Opher, M., Zieger, B., Li, H., Florinski, V., Adhikari, L., Zhu, X., Nakanotani, M.:
Astrophys. J. \textbf{973}, 26 (2024) 

\bibitem[Zimbardo et al.(2006)]{zimb06}
Zimbardo, G., Pommois, P., Veltri, P.: Astrophys. J. \textbf{639}, L91 (2006) 

\bibitem[Zimbardo et al.(2012)]{zimb12}
Zimbardo, G., Perri, S., Pommois, P., Veltri, P.: Adv. Space Res. \textbf{49}, 1633 (2012) 

\end{thebibliography}
\end{document}